\documentclass[letterpaper,twocolumn,10pt]{article}
\usepackage{usenix-2020-09}

\usepackage{amsmath,amsfonts,amsthm}
\usepackage{algorithm,algorithmic} 
\usepackage{xspace}
\usepackage{graphicx}
\usepackage{textcomp}
\usepackage{xcolor}
\usepackage{listings}
\usepackage{multirow}
\usepackage{pifont}
\usepackage{hyperref}
\usepackage{subfig,paralist}
\usepackage{pifont}
\usepackage{enumitem}
\usepackage{url}
\usepackage{booktabs}
\usepackage{tabularx}
\usepackage{multirow,array}
\usepackage{threeparttable}

\usepackage{soul}

\newtheorem*{theorem*}{Theorem}

\usepackage{tcolorbox}
\definecolor{mycolor}{rgb}{0.122, 0.435, 0.698}
\definecolor{gray1}{gray}{0.3}

\definecolor{codegreen}{rgb}{0,0.6,0}
\definecolor{codegray}{rgb}{0.5,0.5,0.5}
\definecolor{codepurple}{rgb}{0.58,0,0.82}
\definecolor{backcolour}{rgb}{0.95,0.95,0.92}
\definecolor{trafogreen}{HTML}{377EB8}
\definecolor{trafopurple}{HTML}{FF7F00}
\definecolor{trafoyellow}{HTML}{4DAF4A}
\definecolor{vptpurple}{HTML}{984EA3}
\definecolor{vptyellow}{HTML}{C4B727}

\DeclareRobustCommand{\hltrafogreen}[1]{{\sethlcolor{trafogreen!30}\hl{#1}}}
\DeclareRobustCommand{\hltrafopurple}[1]{{\sethlcolor{trafopurple!30}\hl{#1}}}
\DeclareRobustCommand{\hltrafoyellow}[1]{{\sethlcolor{trafoyellow!30}\hl{#1}}}
\DeclareRobustCommand{\hlyellow}[1]{{\sethlcolor{vptyellow!30}\hl{#1}}}
\DeclareRobustCommand{\hlpurple}[1]{{\sethlcolor{vptpurple!40}\hl{#1}}}

\lstdefinestyle{mystyle}{
    commentstyle=\color{codegreen},
    keywordstyle=\color{magenta},
    numberstyle=\tiny\color{codegray},
    stringstyle=\color{codepurple},
    basicstyle=\tiny\ttfamily,
    breakatwhitespace=false,
    breaklines=true,
    captionpos=b,
    keepspaces=true,
    numbers=left,
    numbersep=5pt,
    showspaces=false,
    showstringspaces=false,
    showtabs=false,
    tabsize=2,
    columns=fixed
}
\newcommand{\result}[1]{%
\begin{tcolorbox}[colframe=black,boxrule=0.5pt,arc=4pt,
      left=6pt,right=6pt,top=6pt,bottom=6pt,boxsep=0pt,width=\columnwidth]%
      {\emph{#1}}
\end{tcolorbox}%
}

\definecolor{darkgreen}{rgb}{0.0, 0.5, 0.0}
\definecolor{darkred}{rgb}{0.82, 0.1, 0.26}

\hypersetup{
    colorlinks,
    linkcolor={black},
    citecolor={black},
    urlcolor={black}
}

\begin{document}

\date{}

\title{\Large \bf Uncovering the Limits of Machine Learning\\ for Automatic Vulnerability Detection}

\author{
{\rm Niklas Risse}\\
MPI-SP, Germany
\and
{\rm Marcel Böhme}\\
MPI-SP, Germany
} 

\maketitle

\begin{abstract}
Recent results of machine learning for automatic vulnerability detection (ML4VD) have been very promising. Given only the source code of a function $f$, ML4VD techniques can decide if $f$ contains a security flaw with up to 70\% accuracy. However, as evident in our own experiments, the same top-performing models are unable to distinguish between functions that contain a vulnerability and functions where the vulnerability is patched. So, how can we explain this contradiction and how can we improve the way we evaluate ML4VD techniques to get a better picture of their actual capabilities?

In this paper, we identify overfitting to unrelated features and out-of-distribution generalization as two problems, which are not captured by the traditional approach of evaluating ML4VD techniques. As a remedy, we propose a novel benchmarking methodology to help researchers better evaluate the true capabilities and limits of ML4VD techniques. Specifically, we propose (i)~to augment the training and validation dataset according to our cross-validation algorithm, where a semantic preserving transformation is applied during the augmentation of either the training set or the testing set, and (ii)~to augment the testing set with code snippets where the vulnerabilities are patched.

Using six ML4VD techniques and two datasets, we find (a) that state-of-the-art models severely overfit to unrelated features for predicting the vulnerabilities in the testing data, (b) that the performance gained by data augmentation does not generalize beyond the specific augmentations applied during training, and (c) that state-of-the-art ML4VD techniques are unable to distinguish vulnerable functions from their patches.
\end{abstract}

\section{Introduction}
\label{sec:introduction}
Recently several different publications have reported high scores on vulnerability detection benchmarks using machine learning (ML) techniques \cite{codebert_paper, vulberta, cotext, PLBART_paper, graphcodebert_paper, unixcoder_paper}. The resulting models seem to outperform traditional program analysis methods, e.g. static analysis, even without requiring any hard-coded knowledge of program semantics or computational models. So, does this mean that the problem of detecting security vulnerabilities in software is solved? Are these models actually able to detect security vulnerabilities, or do the reported scores provide a false sense of security?

Even though ML4VD techniques achieve high scores on vulnerability detection benchmark datasets, there are still situations in which they fail to meet expectations when presented with new data. For example, it is possible to apply small semantic preserving changes to augment the testing dataset of a state-of-the-art model and then measure whether the model changes its predictions. If it does, it would indicate a dependence of the prediction on unrelated features. Examples of such transformations are identifier renaming \cite{DAMP, AVERLOC, Metropolis_Hastings, ALERT, CARROT}, insertion of unexecuted statements \cite{DAMP, CARROT, AVERLOC, Generating_Adversarial_Computer_Programs_using_Optimized_Obfuscations} or replacement of code elements with equivalent elements \cite{27_transformations, 9678706}. The impact of augmenting testing data using these transformations has been explored for many different software-related tasks and the results seem to be clear: Learning-based models fail to perform well when testing data gets augmented using semantic preserving transformations of code \cite{ALERT, On_the_generalizability, 9678706, AVERLOC, Generating_Adversarial_Computer_Programs_using_Optimized_Obfuscations, DAMP, Abstain_refined_adv_train, Metropolis_Hastings, CARROT}.  

In our own experiments, we were able to reproduce the findings of the literature and made additional observations: ML4VD techniques that were trained on typical training data for vulnerability detection are also unable to distinguish between vulnerable functions and their patched counterparts. If a patched function is also predicted as vulnerable, this indicates that the prediction critically depends on features unrelated to the presence of a security vulnerability.

It has previously been proposed to reduce the dependence on unrelated features by augmenting not just the testing data but also the training data \cite{ALERT, AVERLOC, Generating_Adversarial_Computer_Programs_using_Optimized_Obfuscations, DAMP, Abstain_refined_adv_train, Metropolis_Hastings, CARROT}. Indeed, this seems to restore the lost performance back to previous levels, but does it really reduce the dependence on unrelated features, or are the models just overfitting to different unrelated features of the data?

In this paper, we propose a novel benchmarking methodology that can be used to evaluate the capabilities of ML4VD techniques by using data augmentation. First, we propose \autoref{alg:methodology_1}, in which a selected semantic preserving transformation is applied to the training dataset of a model, and a \emph{different} transformation is applied to the testing dataset. When repeated for all possible pairs out of a set of transformations, the resulting scores provide a better measure of overfitting to the unrelated features that are introduced by the semantic preserving transformations during training data augmentation. Second, we propose \autoref{alg:methodology_2}, in which a trained model is evaluated on a testing dataset that contains both vulnerable programs and their respective patches. The results provide a measure of the model’s ability to generalize to a modified vulnerability detection setting.

In order to validate \autoref{alg:methodology_1} and \autoref{alg:methodology_2} empirically,  we selected six state-of-the-art ML4VD techniques. All evaluated ML4VD techniques happen to be token-based large language models (LLMs). As our selection criterion, we defined the top-performing ML4VD techniques on the most widely known ML vulnerability detection benchmark CodeXGLUE \cite{CODEXGLUE, codexglue_leaderboards} that are available as open source. This gave us ranks 1, 2, 6, 10, and 12 of the leaderboard, all of which are token-based LLMs. In fact, 9 of the Top-10 solutions on the leaderboard are token-based LLMs. By applying \autoref{alg:methodology_1} and \autoref{alg:methodology_2} in our empirical study of six state-of-the-art ML4VD techniques and three datasets, we confirmed that ML4VD techniques continue to leverage unrelated features when deciding whether a function contains a vulnerability.

For \autoref{alg:methodology_1}, we implemented 11 different semantic preserving transformations for data augmentation and evaluated the trained models using two popular vulnerability detection datasets. As expected, we find a strong benefit of training data augmentation (69.0\% and 66.2\% average restoration of accuracy/f1-score for the two datasets) when the transformations applied to training and testing datasets are the same. However, we find no improvement in performance when the transformations applied to training and testing datasets are different. In fact, we even find an additional 30.2\% and 77.5\% average \emph{decrease} in accuracy/f1-score for the two datasets. In other words, ML4VD techniques still severely overfit to the specific label-unrelated features introduced by training data augmentation. The improvement in performance gained by data augmentation only applies to the specific type of transformations used during training.

For \autoref{alg:methodology_2}, we introduce a new dataset, VulnPatchPairs, which contains 26.2k C functions and is derived from the CodeXGLUE/Devign vulnerability detection dataset \cite{Devign}. Exactly half of the functions in VulnPatchPairs contain security vulnerabilities. The other half are patched versions of the first half.\footnote{See \autoref{sec:vuln_patch_pairs} for details.} We investigated six ML4VD techniques using VulnPatchPairs and evaluated their ability to generalize from their typical training data to VulnPatchPairs, and vice versa. To our surprise, all six ML4VD techniques that were trained on a typical training dataset were unable to distinguish between the vulnerable functions and their patched counterparts in VulnPatchPairs. On average, the accuracy turned out to be worse than random guessing. The trained models are unable to generalize from a standard vulnerability detection dataset to the modified setting.
\newline
\noindent
In summary, this paper contributes two novel algorithms that can be used to uncover major problems of ML4VD techniques that are not detected using the standard evaluation setup: Overfitting to semantic preserving code changes and the inability to generalize between related vulnerability detection settings. Additionally, we provide an empirical evaluation of six state-of-the-art ML4VD techniques using the proposed methodology.
\begin{itemize}[leftmargin=*]
    \item We present a general methodology consisting of two algorithms, that can be used to evaluate ML4VD techniques.
    \item We show empirically, that state-of-the-art ML4VD techniques overfit to the unrelated features introduced by semantic preserving transformations during data augmentation.
    \item We introduce \emph{VulnPatchPairs}, a new dataset that contains vulnerable C function and the corresponding patched versions of the same functions. It is available at \newline \url{https://github.com/niklasrisse/VPP}.
    \item We demonstrate, that six state-of-the-art ML4VD techniques are not able to distinguish between the vulnerable and patched functions in VulnPatchPairs.
    \item We publish all of our code and results for reproducibility. They are available at \newline \url{https://github.com/niklasrisse/USENIX_2024}.
\end{itemize}

\section{Related Work}
\label{sec:related_work}
One of the main tools to study the limits of ML4VD techniques are semantic preserving transformations of code. Previous work \cite{ALERT, On_the_generalizability, 9678706, AVERLOC, Generating_Adversarial_Computer_Programs_using_Optimized_Obfuscations, DAMP, Abstain_refined_adv_train, Metropolis_Hastings, CARROT, ZigZag, 27_transformations} proposed methods to generate semantic preserving transformations for source code datasets and investigated their impact when used to augment testing data of learned models. 

Many of the works that reported the failures of learned models when testing data was augmented also investigated training data augmentation using their respective methods \cite{ALERT, AVERLOC, Generating_Adversarial_Computer_Programs_using_Optimized_Obfuscations, DAMP, Abstain_refined_adv_train, Metropolis_Hastings, CARROT}. A common finding in all of these publications is that training data augmentation using a specific type of semantic preserving transformation leads to improved performance on testing sets that have been augmented the same way. But does the performance gained by data augmentation generalize beyond the specific augmentations applied during training?

\begin{figure*}[t]
    \centering
    \subfloat[Code Snippet]{%
    \noindent\begin{minipage}{\linewidth}
\begin{lstlisting}[frame=tb, language=C, linewidth=0.45\linewidth]
static inline int coeff_unpack_golomb(GetBitContext *gb, int qfactor, int qoffset)
{
    int coeff = dirac_get_se_golomb(gb);
    const int sign = FFSIGN(coeff);
    const unsigned sign = FFSIGN(coeff);
    if (coeff)
        coeff = sign*((sign * coeff * qfactor + qoffset) >> 2);
    return coeff;
}
\end{lstlisting}
    \end{minipage}
    }
    \hfill
    \subfloat[Transformed Code Snippet]{%
    \noindent\begin{minipage}{\linewidth}
        \lstset{belowskip=0pt}
\begin{lstlisting}[lastline=7, frame=t, language=C, linewidth=0.45\linewidth]
static inline int coeff_unpack_golomb(GetBitContext *gb, int qfactor, int qoffset)
{
    int coeff = dirac_get_se_golomb(gb);
    const int sign = FFSIGN(coeff);
    const unsigned sign = FFSIGN(coeff);
    if (coeff)
        coeff = sign*((sign * coeff * qfactor + qoffset) >> 2);
\end{lstlisting}
        \lstset{aboveskip=0pt}
\begin{lstlisting}[firstline=8,lastline=9, firstnumber=8, language=C, backgroundcolor=\color{orange!30}, linewidth=0.45\linewidth]
    if (0)
        coeff = 666;
\end{lstlisting}
        \lstset{aboveskip=0pt,belowskip=5pt}
\begin{lstlisting}[firstline=10, firstnumber=10, frame=b, language=C, linewidth=0.45\linewidth]
    return coeff;
}
\end{lstlisting}
    \end{minipage}
    }
    \caption{Example of a simple semantic preserving transformation. The change (orange) has no effect on the vulnerability label. Both code snippets contain a security vulnerability (integer overflow in line 4). The code was taken from the Ffmpeg GitHub repository (URL: \url{https://github.com/FFmpeg/FFmpeg/commit/92da2309}) and is part of the CodeXGLUE/Devign dataset.}
    \label{fig:example_semantic_trafo}
\end{figure*}

Some of the publications that propose methods for data augmentation \cite{AVERLOC, ALERT, Abstain_refined_adv_train, Generating_Adversarial_Computer_Programs_using_Optimized_Obfuscations, CARROT} take it a step further; they augment the training data using a slightly different but related type of transformation than for the testing data. For example, Henkel et al. \cite{AVERLOC} apply their gradient-based approach for identifier renaming to the training data and a random renaming strategy to the testing data. Similarly, Yang et al. \cite{ALERT} apply their method for variable renaming to a training dataset and the method proposed by Zhang et al. \cite{Metropolis_Hastings} to a testing dataset. All of these works find an improved performance when the training dataset is augmented in a similar way than the testing dataset. However, the transformations used for augmenting the training and testing data in these publications are all similar in type, e.g. they both rename identifiers. But does the performance also improve when training data is augmented in a different way than the testing data? Our work aims to fill this gap in the literature by carrying out a thorough empirical study that considers a diverse set of 11 different transformations, six state-of-the-art ML4VD techniques, and two high-quality datasets.

Similar to other related publications listed above \cite{ALERT, AVERLOC, Generating_Adversarial_Computer_Programs_using_Optimized_Obfuscations}, Rahman et al. \cite{towards_causal} investigate overfitting of ML4VD techniques to variable and API names by transforming them in the testing data. Additionally, they propose a new method to address the overfitting based on causal learning, which aims to disable models from using superficial features (e.g. variable names) entirely. While their approach seems to be effective to avoid overfitting for concrete and simple transformations (e.g. changing variable names), the authors do not investigate how their method performs when faced with unseen semantic preserving transformations, that were not specifically trained for. Our proposed methodology can be used as a tool to do this, which allows to draw conclusions about overfitting to unrelated features fundamentally, irrespective of the type of transformation applied to training- and testing data.

In order to evaluate the general capabilities of ML4VD techniques, we collected a new dataset (VulnPatchPairs), which contains both vulnerable functions and their respective patches. The collection of a pairwise vulnerability-patch dataset has been proposed by previous work \cite{Fix_dataset_1, fix_dataset_2, fix_dataset_3}, e.g. for the research field of automated fixing. However, to the best of our knowledge, we are the first to utilize such a dataset to evaluate the general capabilities of ML4VD techniques.

Two recently published papers \cite{diversevul, empirical_study} report poor generalization capabilities of different ML-based techniques (e.g. LLMs and GNNs) when evaluated on functions from unseen git projects. Our \autoref{alg:methodology_2} also investigates the generalization capabilities of ML4VD techniques, but using a different setup, in which functions in the evaluation data belong to a modified vulnerability detection setting (e.g. vulnerable functions and their patches), but can be from the same projects.

\section{Methodology}
\label{sec:methodology}
We propose a novel benchmarking methodology to help researchers better evaluate advances in ML4VD techniques. The methodology consists of two parts, \autoref{alg:methodology_1} (A1) and \autoref{alg:methodology_2} (A2).

\subsection{Data Augmentation}
\label{sec:transformations_definition}

A central component of our methodology is data augmentation, and the expectations for vulnerability detection models that emerge from using code transformations for data augmentation.

We define data augmentation as the application of one or multiple code transformations onto all code snippets of a given code snippet dataset $CD \subset \mathcal{C}$, where $\mathcal{C}$ is a space that represents all possible code snippets $c \in \mathcal{C}$ in a given programming language.

A code transformation $t: \mathcal{C} \to \mathcal{C}$ is a function that maps from and to $\mathcal{C}$. Let's assume we have an oracle function $g : \mathcal{C} \to \{ 0, 1 \}$, which maps from the space of code snippets $\mathcal{C}$ to either 0 or 1. The oracle function $g$ represents the ground truth, i.e. it shows whether a code snippet $c$ does (1) or does not (0) contain a security vulnerability. For a given code snippet dataset $CD \subset \mathcal{C}$, a code transformation $t$ can be characterized by its effect on $g(t(c))$ $\forall$ $ c \in CD$:

\textbf{Semantic Preserving Transformation.} We call a transformation $t_p$ semantic preserving w.r.t. $CD$, if the changes introduced by applying it do not affect the ground truth vulnerability label, $g(c) = g(t_p(c))$ $\forall$ $ c \in CD$. \autoref{fig:example_semantic_trafo} shows an example of a simple semantic preserving transformation applied to a real-world code snippet. 

\textbf{Label Inverting Transformation.} We call a transformation $t_d$ label inverting w.r.t. CD, if the changes introduced by applying it change the ground truth vulnerability label, $g(c) \neq g(t_d(c))$ $\forall$ $ c \in CD$. In other words, a label inverting transformation either adds or removes a vulnerability from a code snippet.

In general, we expect a vulnerability detection model to correctly predict, whether a given code snippet contains a security vulnerability, independent of any semantic preserving or label inverting transformations that have been previously applied to the code snippet. Specifically, we can formulate the following expectations:
\begin{enumerate}
    \item If we change a code snippet without affecting the vulnerability label (semantic preserving transformation), we expect a vulnerability detection tool to compute the same correct prediction as before applying the change. 
    \item If we add or remove a vulnerability from a code snippet (label inverting transformation), we expect a vulnerability detection tool to still deliver a correct prediction, or i.e. we expect it to change its prediction with the ground truth label of the code snippet.
\end{enumerate} 
In the following sections, we present two algorithms, which allow to evaluate ML4VD techniques using the two formulated expectations.

\subsection{A1: Detecting Overfitting to Code Changes}

\begin{figure}[t]
    \centering
    \includegraphics[width=0.75\linewidth]{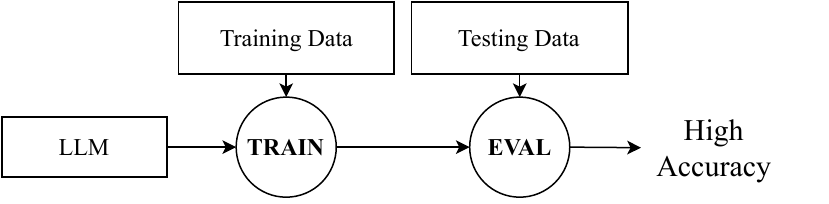} \\
    \vspace{0.15cm}
    \includegraphics[width=0.75\linewidth]{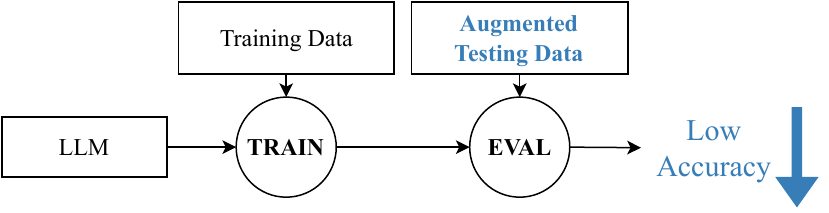} \\
    \vspace{0.15cm}
    \includegraphics[width=0.75\linewidth]{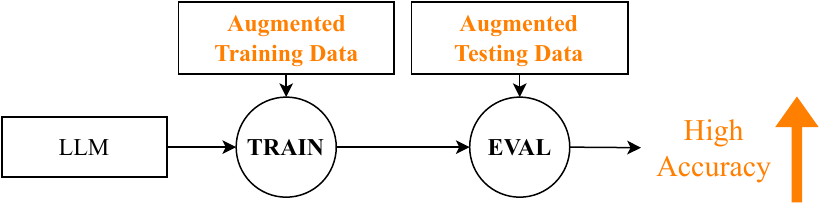} \\
    \vspace{0.15cm}
    \includegraphics[width=0.75\linewidth]{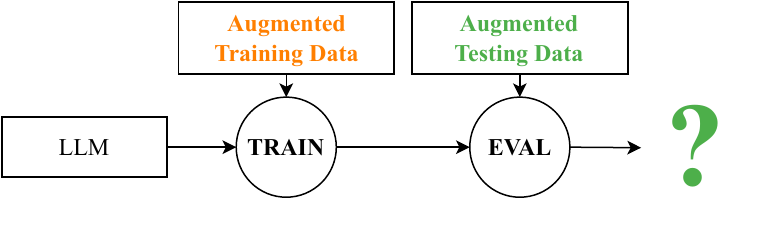}
    \caption{Visualization of \autoref{alg:methodology_1}, which we created to detect overfitting of ML4VD techniques to unrelated features introduced by data augmentation. Colors represent that either only testing data is augmented (blue), training- and testing data are augmented using the same (orange), or different augmentation methods (green).}
    \label{fig:story_1}
\end{figure}

\begin{algorithm}[t]
\caption{Detecting Overfitting to Code Changes}
\label{alg:methodology_1}
\begin{algorithmic}[1]
\scriptsize
\REQUIRE {Semantic Preserving Transformations $T := \{ t_1, ..., t_N\}$ \newline Training Dataset $Tr$ \newline Testing Dataset $Te$ \newline ML Training Method \emph{train\_model} \newline ML Evaluation Method \emph{evaluate\_model} \newline Performance Metric $M$}
\STATE {$MLM[Tr]=$\emph{ train\_model}$(Tr)$}
\STATE {$score[MLM[Tr], Te]=$\emph{ evaluate\_model}$(MLM[Tr], Te, M)$}
\FOR {\textbf{each} $t_k\in T$}
\STATE $Te_k = t_k(Te)$ \quad\quad\textcolor{gray}{\emph{// testing data augmentation}}
\STATE {\hltrafogreen{$score[MLM[Tr], Te_k]=$\emph{ evaluate\_model}$(MLM[Tr], Te_k, M)$}}
\STATE {\hltrafogreen{$effect[Tr, Te_k] = score[MLM[Tr], Te_k] - score[MLM[Tr], Te]$}}
\STATE $Tr_k = t_k(Tr)$ \quad\quad\textcolor{gray}{\emph{// training data augmentation}}
\STATE {$MLM[Tr_k]=$\emph{ train\_model}$(Tr_k)$}
\STATE {\hltrafopurple{$score[MLM[Tr_k], Te_k]=$\emph{ evaluate\_model}$(MLM[Tr_k], Te_k, M)$}}
\STATE {\hltrafopurple{$MLM[Tr_k, Te_k] = score[MLM[Tr_k], Te_k] - score[MLM[Tr], Te]$}}
\FOR {\textbf{each} $t_{j \neq k} \in T$}
\STATE $Te_j = t_j(Te)$ \quad\textcolor{gray}{\emph{// testing data augmentation}}
\STATE {\hltrafoyellow{$score[MLM[Tr_k], Te_j]=$\emph{ evaluate\_model}$(MLM[Tr_k], Te_j, M)$}}
\STATE {\hltrafoyellow{$effect[Tr_k, Te_j] = score[MLM[Tr_k], Te_j] - score[MLM[Tr], Te]$}}
\ENDFOR
\ENDFOR
\ENSURE {\hltrafogreen{$output_{A1.1} = (\sum_k effect[Tr, Te_k]) / N$} \newline 
\hltrafopurple{$output_{A1.2} = (\sum_k effect[Tr_k, Te_k]) / N$} \newline 
\hltrafoyellow{$output_{A1.3} = (\sum_k \sum_{j \neq k} effect[Tr_k, Te_j]) / (N(N-1))$ }}
\end{algorithmic}
\end{algorithm}

The goal of \autoref{alg:methodology_1} is to measure, whether ML4VD techniques overfit to augmentations of their training data that are unrelated to the respective vulnerability labels and whether the performance gained by data augmentation generalizes beyond the specific augmentations applied during training. We provide a simple visualization of the idea behind the algorithm in \autoref{fig:story_1}, the algorithm itself in \autoref{alg:methodology_1}, and a description of the most important parts in the following paragraphs. We use the colors \hltrafogreen{blue}, \hltrafopurple{orange}, and \hltrafoyellow{green}, to connect the basic ideas of the algorithm with the experimental results across the paper\footnote{See \autoref{fig:story_1}, \autoref{alg:methodology_1}, \autoref{fig:rq_1}, \autoref{fig:rq_2}, \autoref{fig:rq_2:c} and \autoref{tab:results_alg_1}.}. 

\textbf{What are the inputs?} The inputs of \autoref{alg:methodology_1} are a set of different semantic preserving transformations $T := \{ t_1, ..., t_N\}$, a training dataset $Tr$, a testing dataset $Te$, a ML training method \emph{train\_model}, a ML evaluation method \emph{evaluate\_model}, and a performance metric M. The training and testing datasets Tr and Te consist of code-label pairs $(c_i, v_i)$, with $c_i \in \mathcal{C}$ representing code snippets and $v_i \in \{ 0, 1 \}$ representing labels that indicate the absence (0) or presence (1) of security vulnerabilities in the respective code snippets. The method \emph{train\_model} can utilize the training dataset Tr to train a machine learning model $MLM : \mathcal{C} \to \{ 0, 1 \}$, which maps from the space of code snippets $\mathcal{C}$ to either 1 (vulnerability) or 0 (no vulnerability). The method \emph{evaluate\_model} can use the performance metric M to quantify and aggregate the performance of a trained model $MLM$ on a testing dataset Te into a single number between 0 (bad) and 1 (perfect).

\textbf{What is computed?} \autoref{alg:methodology_1} computes the average effects of (a) augmenting the testing data of the selected ML technique using transformations $t_k \in T$ ($output_{A1.1}$), (b) using the same transformations to also augment the training data ($output_{A1.2}$), and (c) using different transformations to also augment the training data ($output_{A1.3}$).

In lines 4-6, Algorithm 1 computes the effect of augmenting the testing dataset Te with the transformation $t_k$ on the performance of the trained model $MLM[Tr]$. The result is $effect[Tr, Te_k]$, the absolute difference between the scores of $MLM[Tr]$ on the clean testing dataset $Te$ and the augmented testing dataset $Te_k$. In other words, $effect[Tr, Te_k]$ quantifies how many points in score are lost if we augment the testing dataset with transformation $t_k$. $output_{A1.1}$ aggregates this intermediate result over all transformations $t_k \in T$.

\begin{figure}[t]
    \centering
    \includegraphics[width=0.75\linewidth]{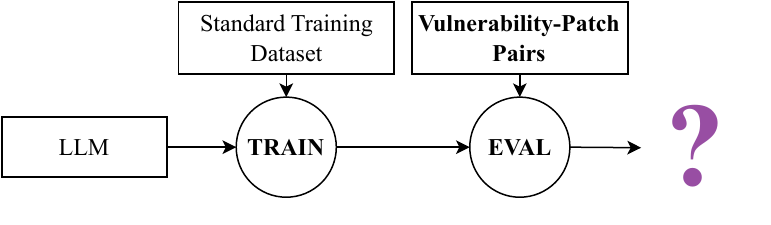} \\
    \vspace{0.15cm}
    \includegraphics[width=0.75\linewidth]{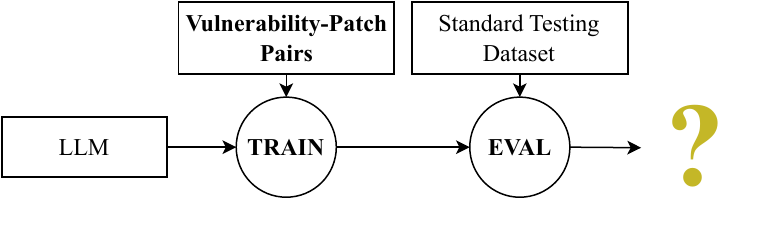}
    \caption{Visualization of \autoref{alg:methodology_2}, which we created to test whether ML4VD techniques are able to generalize to a modified setting, which requires to distinguish between vulnerabilities and patches.}
    \label{fig:story_2_2}
\end{figure}

In lines 7-10, Algorithm 1 goes a step further and computes the effect of both augmenting the training dataset $Tr$ and the testing dataset $Te$ using the same transformation $t_k$. The result is $effect[Tr_k, Te_k]$, the absolute difference between scores of $MLM[Tr_k]$ on the augmented testing dataset $Te_k$ and $MLM[Tr]$ on the testing dataset $Te$. In other words, $effect[Tr_k, Te_k]$ quantifies how many points in score are lost if we augment both the training and the testing dataset with transformation $t_k$. $output_{A1.2}$ aggregates this intermediate result over all transformations $t_k \in T$.

In lines 12-14, the algorithm computes the effect of augmenting the testing dataset using a \emph{different} transformation $t_j$ than for the training dataset. The result is $effect[Tr_k, Te_j]$, the absolute difference between scores of $MLM[Tr]$ on the testing dataset $Te$ and $MLM[Tr_k]$ on the augmented testing $Te_j$. In other words, $effect[Tr_k, Te_j]$ quantifies how many points in score are lost if we augment the training and the testing dataset with different transformations $t_k$ and $t_j$. $output_{A1.3}$ aggregates this intermediate result over all transformations $t_k \in T, t_{j \neq k} \in T$.

\textbf{How can the results be used?} Using this algorithm, researchers can effectively evaluate new ML4VD techniques. Specifically, for a selected technique researchers can answer the following questions:
\begin{enumerate}
    \item How much does the performance of the selected ML technique decrease if we augment the code snippets for testing without affecting the vulnerability labels? Answer: On average, the performance does change by $output_{A1.1}$ points.
    \item How much performance of the selected ML technique can be restored, if we augment the training code snippets in a similar way than the testing code snippets? Answer: On average, $output_{A1.2} - output_{A1.1}$ of the initial decrease can be restored. 
    \item How much performance of the selected ML technique can be restored, if we augment the training code snippets in a different way than the testing code snippets? Answer: On average, $output_{A1.3} - output_{A1.1}$ of the initial decrease can be restored. 
    \item Does the selected ML technique overfit to specific augmentations of the training data that are unrelated to the respective vulnerability labels? Answer: If $output_{A1.2} >> output_{A1.3}$: Yes, otherwise No.
\end{enumerate}

\begin{algorithm}[t]
\caption{Distinguish between Vulnerability and Patch}
\label{alg:methodology_2}
\begin{algorithmic}[1]
\scriptsize
\REQUIRE {Standard Training Dataset Tr \newline Standard Testing Dataset Te \newline Vulnerability-Patch Training Dataset VPTr \newline Vulnerability-Patch Testing Dataset VPTe \newline ML Training Method \emph{train\_model} \newline ML Evaluation Method \emph{evaluate\_model} \newline Performance Metric M}
\STATE {$MLM[Tr]=$\emph{ train\_model}$(Tr)$}
\STATE {$MLM[VPTr]=$\emph{ train\_model}$(VPTr)$}
\STATE {$score[MLM[Tr], Te]=$\emph{ evaluate\_model}$(MLM[Tr], Te, M)$}
\STATE {\hlpurple{$score[MLM[Tr], VPTe]=$\emph{ evaluate\_model}$(MLM[Tr], VPTe, M)$}}
\STATE {$score[MLM[VPTr], VPTe]=$\emph{ evaluate\_model}$(MLM[VPTr], VPTe, M)$}
\STATE {\hlyellow{$score[MLM[VPTr], Te]=$\emph{ evaluate\_model}$(MLM[VPTr], Te, M)$}}
\ENSURE {$output_{A2.1} = score[MLM[Tr], Te]$} \newline
\hlpurple{$output_{A2.2} = score[MLM[Tr], VPTe]$} \newline
$output_{A2.3} = score[MLM[VPTr], VPTe]$ \newline 
\hlyellow{$output_{A2.4} = score[MLM[VPTr], Te]$}
\end{algorithmic}
\end{algorithm}

\subsection{A2: Distinguish between Vulnerability and Patch}
The main goal of \autoref{alg:methodology_2} is to evaluate, whether ML4VD techniques are able to generalize from their typical training data to a modified setting, which requires to distinguish security vulnerabilities from their patches. Additionally, the algorithm also aims to evaluate the reverse, or i.e. whether ML4VD techniques that were trained to distinguish between vulnerabilities and their patches are able to perform well on standard testing data. We provide a simple visualization of the idea behind the algorithm in \autoref{fig:story_2_2}, the algorithm itself in \autoref{alg:methodology_2}, and a description of the most important parts in the following paragraphs. We use the colors \hlpurple{purple} and \hlyellow{yellow} to connect the basic ideas of the algorithm with the experimental results across the paper \footnote{See \autoref{fig:story_2_2}, \autoref{alg:methodology_2}, and \autoref{tab:results_alg_2}.}.

\textbf{What are the inputs?} In addition to the inputs of \autoref{alg:methodology_1}, \autoref{alg:methodology_2} requires a special vulnerability-patch testing dataset $VPTe$ and a vulnerability-patch training dataset $VPTr$. $VPTe$ and $VPTr$ also consist of code snippets $c_i \in \mathcal{C}$ and vulnerability labels $v_i \in \{ 0, 1\}$, but for every code snippet $c_j$ with label $v_j = 0$, they also contain a snippet $c_{l \neq j}$ with $v_l = 1$ which represents the patched version of $c_j$. The relationship between a code snippet and its patched version can be characterized as a label inverting transformation $t_{PATCH}: \mathcal{C} \to \mathcal{C}$, which maps code snippets $c_j$ to their patched versions $c_l$.

\textbf{What is computed?} The purpose of \autoref{alg:methodology_2} is to quantify the ability of the selected ML technique to generalize between two related vulnerability detection settings. The first setting, represented by the standard training and testing datasets $Tr$ and $Te$, consists of code snippets, which either contain or do not contain a vulnerability. This setting is most frequently used in the related literature \cite{unixcoder_paper, graphcodebert_paper, codebert_paper, vulberta, cotext, PLBART_paper}. The second setting, represented by the vulnerability-patch training and testing datasets $VPTr$ and $VPTe$, consists of vulnerable code snippets and their respective patches. As formulated in \autoref{sec:transformations_definition}, a perfect vulnerability detection model should be able to perform well in both settings, irrespective of the setting on which it was trained. In other words, a vulnerability detection model should be able to generalize between the settings.

In total, \autoref{alg:methodology_2} computes four scores. In line 3, \autoref{alg:methodology_2} computes the score of a model $MLM[Tr]$, which was trained on the standard training dataset $Tr$, on the standard testing dataset $Te$. This score represents the standard evaluation and serves as a baseline for the other scores.

In line 4, \autoref{alg:methodology_2} computes the score of a model $MLM[Tr]$, which was trained on the standard training dataset $Tr$, on the vulnerability-patch testing dataset $VPTe$. When compared to the first score, this result serves as a measure of $MLM[Tr]$s ability to generalize to the modified setting, which requires to distinguish vulnerabilities from their patches.

In line 5, \autoref{alg:methodology_2} computes the score of a model $MLM[VPTr]$, which was trained on the vulnerability-patch training dataset $VPTr$, on the vulnerability-patch testing dataset $VPTe$. Again, this score serves as a baseline for the other scores.

In line 6, \autoref{alg:methodology_2} computes the score of a model $MLM[VPTr]$, which was trained on the vulnerability-patch training dataset $VPTr$, on the standard testing dataset $Te$. When compared to the third score, this result serves as a measure of $MLM[VPTr]$s ability to generalize back to the standard vulnerability detection setting. 

The four computed scores are returned as the outputs of the algorithm ($output_{A2.1}$, $output_{A2.2}$, $output_{A2.3}$ and $output_{A2.4}$).

\textbf{How can the results be used?} Using \autoref{alg:methodology_2}, researchers can effectively evaluate whether the high scores of ML4VD techniques are specific to the testing datasets on which they were computed. Specifically, for a selected technique researchers can answer the following questions:

\begin{enumerate}
    \item Does the performance of the selected ML technique generalize from a standard vulnerability detection dataset to a modified setting, which requires to distinguish vulnerabilities from their patches? Answer: The selected ML technique can distinguish between vulnerabilities and their patches with performance $output_{A2.2}$, compared to a score of $output_{A2.1}$ on the standard testing dataset.
    \item Does the performance of the selected ML technique generalize back to the standard vulnerability detection setting when it is explicitly trained to distinguish vulnerabilities from their patches? Answer: The selected ML technique achieves a score of $output_{A2.4}$ on the standard testing dataset, compared to a score of $output_{A2.3}$ in the modified setting.
\end{enumerate}

\section{Experimental Setup}
\label{sec:experimental_setup}
\subsection{Research Questions}
\label{sec:research_questions}
Our objective is to validate empirically, whether the two proposed algorithms can be used to evaluate state-of-the-art ML4VD techniques. Specifically, we aim to answer the following research questions.

\noindent
\hangindent=2em
\hangafter=1
\textbf{RQ.1 How is the performance of ML4VD techniques affected, if we augment the input code snippets without affecting the vulnerability labels?} (a) Can we measure a decrease in performance, if we augment the testing data of ML4VD techniques using semantic preserving transformations? (b) Does training data augmentation using the same transformations restore the initial performance? (c) Are there differences between the individual transformations?

\noindent
\hangindent=2em
\hangafter=1
\textbf{RQ.2 Do ML4VD techniques overfit to specific augmentations of their training data that do not affect the respective vulnerability labels?} Can we still restore the performance, if we augment the training dataset with a different semantic preserving transformation than the testing dataset?

\noindent
\hangindent=2em
\hangafter=1
\textbf{RQ.3 Are the high scores of ML4VD techniques specific to benchmark datasets or do they generalize to a modified vulnerability detection setting?} (a) Are state-of-the-art ML4VD techniques able to distinguish between vulnerable functions and their patches? (b) Does training to distinguish between vulnerable functions and their patches improve the performance on standard testing data?

\subsection{Semantic Preserving Transformations}
\label{sec:semantic_preserving_transformations}
One of the central components of algorithms \ref{alg:methodology_1} and \ref{alg:methodology_2} are semantic preserving transformations of code. The most common semantic preserving transformations that are used in the literature to investigate the limits of learned models for source code related tasks are identifier renaming \cite{DAMP, AVERLOC, Metropolis_Hastings, Generating_Adversarial_Computer_Programs_using_Optimized_Obfuscations, ALERT, CARROT}, insertion of unexecuted statements \cite{DAMP, CARROT, AVERLOC, Generating_Adversarial_Computer_Programs_using_Optimized_Obfuscations}, replacement of statements with equivalent statements \cite{27_transformations}, reordering of unrelated statements \cite{orvalho2022multipas}, deletion of unexecuted statements (e.g. comments) \cite{27_transformations}, or combinations of the before mentioned \cite{AVERLOC, Generating_Adversarial_Computer_Programs_using_Optimized_Obfuscations, CARROT}.

\begin{table}[t]
\centering
\scriptsize
  \caption{The semantic preserving transformations that we used in our experiments.}
  \label{tab:transformations}
  \begin{tabular}{c c p{4cm}}
    \toprule
    Identifier & Type & Description\\
    \midrule
    $t_1$ & Identifier Renaming &  Rename all function parameters to a random token.\\
    $t_2$ & Statement Reordering &  Reorder all function parameters.\\
    $t_3$ & Identifier Renaming &  Rename the function.\\
    $t_4$ & Statement Insertion &  Insert unexecuted code.\\
    $t_5$ & Statement Insertion &  Insert comment.\\
    $t_6$ & Statement Reordering &  Move the code of the function into a separate function.\\
    $t_7$ & Statement Insertion &  Insert white space.\\
    $t_8$ & Statement Insertion &  Define additional void function and call it from the function.\\
    $t_9$ & Statement Removal &  Remove all comments.\\
    $t_{10}$ & Statement Insertion &  Add code from training set as comment.\\
    $t_{11}$ & Random Transformation &  One transformation sampled from $\{ t_1, \ldots, t_{10} \}$ is applied to each function.\\
    \bottomrule
  \end{tabular}
\end{table}

\autoref{tab:transformations} shows the 11 semantic preserving transformations that we implemented for the experiments presented in this paper. We tried to cover all types of transformations commonly used in the literature. The table lists all transformations, categorizes them by type, and provides short descriptions for each of them.

Since ML4VD techniques are ultimately aimed at detecting security vulnerabilities in real-world code, augmenting code using our semantic preserving transformations should also result in code that looks natural, or i.e. looks like it could occur in real-world software. To address this, we experimentally confirmed that our semantic preserving transformations do not decrease the naturalness (measured by cross entropy) of the code. We provide more information on this as supplementary material in \autoref{appendix:naturalness}.
\subsection{Vulnerability Detection Datasets}
\label{sec:datasets}
We use two publicly available vulnerability detection datasets for our experiments.

\textbf{CodeXGLUE/Devign}. CodeXGLUE is a machine learning benchmark for code understanding and generation \cite{CODEXGLUE}. It consists of several datasets for different source code related tasks. In our experiments, we only use the dataset for vulnerability detection, which is based on the Devign dataset \cite{Devign}. Throughout this paper, we refer to this dataset as the \emph{CodeXGLUE/Devign dataset} or just as the CodeXGLUE dataset. The CodeXGLUE dataset contains 26.4k C functions, from which 45.6\% contain vulnerabilities, i.e. the dataset is fairly balanced. The types of vulnerabilities were not formally classified, but based on the collection process the authors found most vulnerabilities in the dataset to be memory-related, e.g. memory leaks, buffer overflows, memory corruption, or crashes. The authors of the CodeXGLUE benchmark also maintain a leaderboard \cite{codexglue_leaderboards}, which tracks the performance of popular learning-based techniques on the different datasets of the benchmark. 

\textbf{VulDeePecker}. The other vulnerability detection dataset that we use in this paper is the Code Gadget Database, which was introduced with the VulDeePecker bug detector \cite{VulDeePecker}. We refer to this dataset as the \emph{VulDeePecker dataset}. The original dataset contains 61.6k C/C++ code samples derived from the NVD \cite{NVD_database} and the SARD \cite{SARD_dataset}, from which 17.7k contain vulnerabilities, mainly buffer (CWE-119) and resource management errors (CWE-399).

\subsection{New Dataset: VulnPatchPairs}
\label{sec:vuln_patch_pairs}
\begin{figure}[t]
    \centering
    \includegraphics[width=\linewidth]{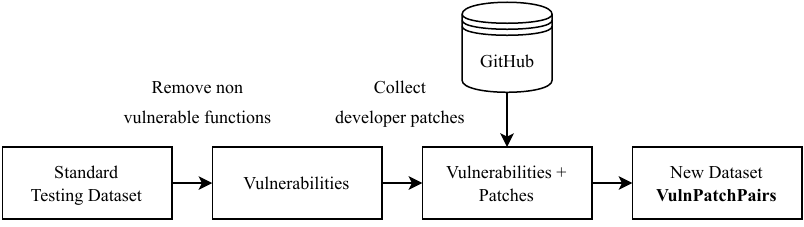}
    \caption{Visualization of the collection process for our new dataset VulnPatchPairs.}
    \label{fig:story_2_1}
\end{figure}

In order to investigate the ability of ML4VD techniques to distinguish between vulnerabilities and their patches (RQ3), we collected a new dataset, which we call \emph{VulnPatchPairs}. We provide a simple visualization of the collection process for VulnPatchPairs in \autoref{fig:story_2_1}.

VulnPatchPairs is an extension of the CodeXGLUE/Devign vulnerability detection dataset \cite{Devign}, which consists of C functions from two popular open source repositories, FFmpeg\footnote{FFmpeg repository: \url{https://github.com/FFmpeg/FFmpeg}} and Qemu\footnote{Qemu repository: \url{https://github.com/qemu/qemu}}. The creators of the dataset describe the collection process in their original publication \cite{Devign}. As a first step, they filtered the selected repositories for security-related commits using a list of keywords. Then, they invested 600 work hours of a four-person team of security researchers to classify the security-related commits into vulnerability-fix commits (VFCs) and non vulnerability-fix commits (non-VFCs) and extracted the respective functions before the commits were applied as vulnerable (before VFCs) and non-vulnerable (before non VFCs) functions. The actual patched versions of the functions (after the VFCs were applied) are not part of their original dataset. However, for each function of their dataset, the authors released the respective commit ID from the two open-source repositories as additional information. We used this information to extract the actual patched versions of the vulnerable functions in the CodeXGLUE dataset from the FFmpeg and Qemu repositories and created a new dataset: VulnPatchPairs.

We manually verified 100 randomly chosen functions from the CodeXGLUE dataset that were labelled as vulnerable. We found 68 out of these 100 functions to actually contain a security vulnerability, 23 to contain no security vulnerability, and 9 with no decision after 10 minutes of manual effort. The results are available at \url{https://github.com/niklasrisse/VPP_label_accuracy}. Even though these results show that there are inaccurate labels in the CodeXGLUE dataset, we do not expect any changes to our main findings. While the absolute performance might change if we used perfectly labeled data, we expect the relative performance (e.g., augmented vs non-augmented) to remain comparable for all of our individual evaluations.

In total, VulnPatchPairs consists of 26.2k C functions from the two open source repositories FFmpeg and Qemu. Exactly half (13.1k) of the 26.2k functions contain security vulnerabilities and were copied from the CodeXGLUE/Devign vulnerability detection dataset. The other 13.1k are the respective patches of the vulnerable functions, which we extracted from the open-source repositories. We published VulnPatchPairs as supplementary material in an open GitHub repository\footnote{VulnPatchPairs: \url{https://github.com/niklasrisse/VPP}}.

\subsection{Machine Learning Techniques}
\label{sec:ml_techniques}

We selected six state-of-the-art ML4VD techniques for our experiments.

\textbf{Selection Criteria.} In order to select techniques that represent the state-of-the-art of machine learning for vulnerability detection, we chose the top techniques from the CodeXGLUE leaderboard \cite{codexglue_leaderboards} for which the authors provide open-source implementations. Measured by citations\footnote{\url{https://api.semanticscholar.org/CorpusID:231855531}} (496) and GitHub Stars\footnote{\url{https://github.com/microsoft/CodeXGLUE}} (1.2k), CodeXGLUE is the most well-known benchmark for source code related machine learning techniques. The vulnerability detection dataset of the benchmark \cite{Devign} is also highly cited (407 citations\footnote{\url{https://api.semanticscholar.org/CorpusID:202539112}}) and widely used to evaluate ML4VD techniques for automatic vulnerability detection.

\textbf{Selected Techniques.} Based on the described criteria, we selected UniXcoder \cite{unixcoder_paper}, CoTexT \cite{cotext}, VulBERTa \cite{vulberta}, PLBart \cite{PLBART_paper}, and CodeBERT \cite{codebert_paper} for our experiments. At submission time of this paper, the six techniques hold rank 1 (UniXcoder), rank 2 (CoTexT), rank 6 (VulBERTa), rank 10 (PLBart), and rank 12 (CodeBERT) on the CodeXGLUE leaderboard for vulnerability detection \cite{codexglue_leaderboards}. In addition to the five techniques from the CodeXGLUE leaderboard, we selected GraphCodeBERT (abbreviated as \emph{GraphCB} in \autoref{tab:results_alg_2}) \cite{graphcodebert_paper}, a technique related to CodeBERT, which utilizes graph representations of source code during the training process.

\textbf{Model Details.} Since the selected techniques belong to the same family of techniques, they also share the same basic architecture:
\begin{enumerate}
    \item Tokenization: A given code function is split into tokens (small sequence of characters that forms a semantic entity), based on a given strategy.
    \item Embedding: Tokens are transformed into numbers, usually by indexing via a learned vocabulary and the addition of positional information.
    \item Transformer Network: Several steps of parametrized computation are applied, resulting in a final embedding.
    \item Prediction Layer: The final layer of the model is a parametrized prediction layer, which computes an output number between 0 and 1 based on the final embedding of the previous step. The output number represents the predicted probability that the input function contains a security vulnerability.
\end{enumerate}
During model training, the parametrized computational layers of steps 3. and 4. are optimized to compute the correct predictions for given training data. The six selected techniques mainly differ in tokenization strategy, training data, optimization objective, and the specific transformer network architecture used. We provide additional information on the specific models in \autoref{appendix:model_details}.

The authors of all six techniques provide publicly available implementations of their techniques \cite{unixcoder_paper, vulberta, cotext, PLBART_paper, codebert_paper, graphcodebert_paper}.

\subsection{Model Training Pipeline}
\label{sec:model_training_pipeline}
 
We used a similar training setup for all model instances that we trained for our experiments.

\textbf{Pre-training.} All models that we train in our experiments have been pre-trained by the authors of the respective publications using various source code datasets. The size of the pre-training datasets spans from 2.3 million (VulBERTa) to 680 million code snippets (PLBart). The original publications provide more information on the pre-training datasets \cite{cotext, vulberta, PLBART_paper, unixcoder_paper}. We use the pre-trained models released by the authors of the respective techniques as starting points for our pipeline and finetune the models on our selected datasets.

\textbf{Data split.} For the CodeXGLUE/Devign dataset, we used the train-/validation-/testing dataset split provided by the authors of the benchmark \cite{CODEXGLUE}, which resulted in a training dataset with 21k functions, a validation dataset with 2.7k functions, and a testing dataset with 2.7k functions. For the VulDeePecker dataset, we used the split provided by Hanif et al. \cite{vulberta}, which resulted in a training dataset with 128.1k functions, a validation dataset with 16k functions, and a testing dataset with 16k functions. The split for VulnPatchPairs is derived from the split of CodeXGLUE, such that \emph{all and only} vulnerable functions in training, validation, and testing sets, respectively, of CodeXGLUE were taken as training, validation, and testing sets of VulnPatchPairs, augmented by their corresponding patches.

\begin{figure*}[t]
    \subfloat[Test set accuracy over ten training epochs of different models trained with VulBERTa on the CodeXGLUE/Devign dataset. Augmenting the testing set $Te$ with transformation $t_{10}$ (blue) decreases the accuracy, but applying the same transformation also to the training dataset $Tr$ (orange) restores the accuracy back to previous levels.\label{fig:rq_1:a}]{%
      \includegraphics[width=0.47\linewidth]{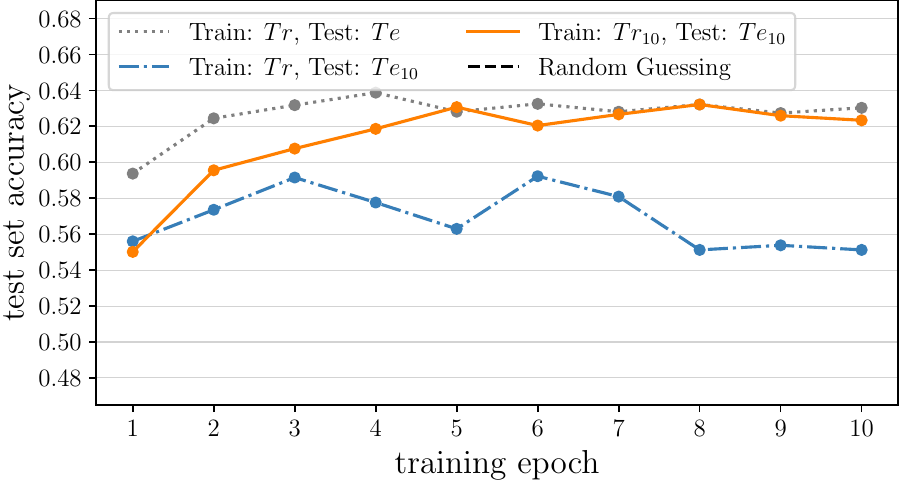}
    }
    \hfill
    \subfloat[Extension of the results in \autoref{fig:rq_1:a}, for all transformations $t_k \in T$, and for all six ML4VD techniques. The boxplots represent distributions of the resulting accuracies. Augmenting the testing set $Te$ with transformations $t_k \in T$ (blue boxplots) decreases the accuracy, but applying the same transformation also to the training dataset $Tr$ (orange boxplots) partially restores the accuracy, although not always to its previous levels.  \label{fig:rq_1:b}]{%
      \includegraphics[width=0.47\linewidth]{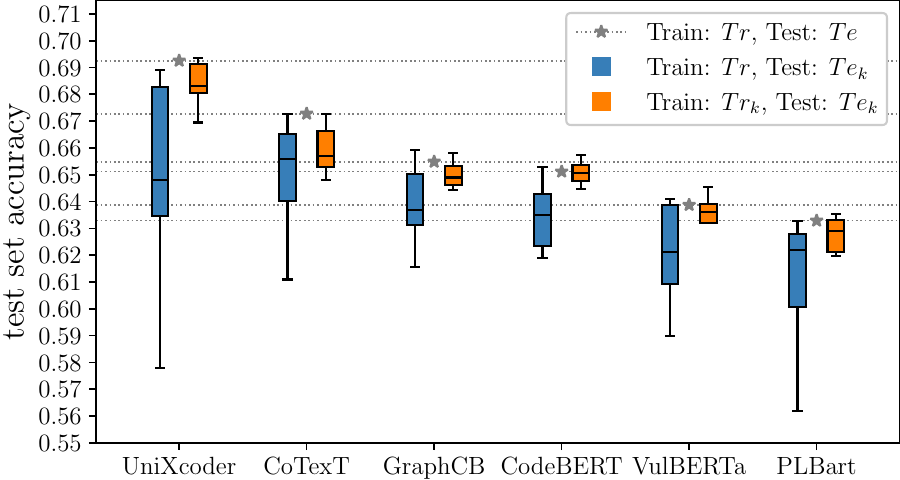}
    }
    \caption{Effects of augmenting the testing data and the training data using the same semantic preserving transformations.}
    \label{fig:rq_1}
\end{figure*}
\begin{figure*}[t]
    \centering
    \includegraphics[width=\linewidth]{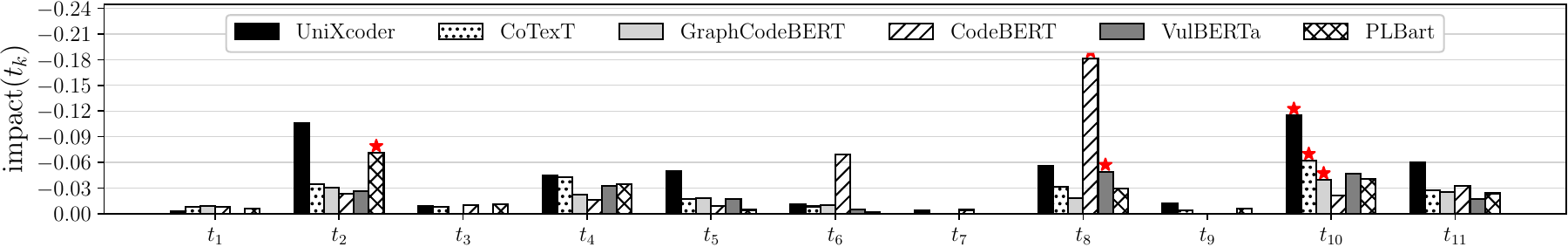}
    \caption{Impact on accuracy caused by augmenting the testing data using the individual transformations $t_k \in T$ ($impact(t_k) := accuracy[MLM[Tr], Te_k] - accuracy[MLM[Tr], Te]$). The most impactful transformations for each ML technique are marked by red stars.}
    \label{fig:rq_2_barchart}
\end{figure*}

\textbf{Pre-processing.} For the VulDeePecker dataset, we removed all duplicate functions and also replaced all label-revealing tokens (e.g. comment with token "bad" above a vulnerable function) that we found by manual inspection of the dataset with randomly selected tokens. For the CodeXGLUE and VulnPatchPairs datasets, we applied no additional pre-processing steps.

\textbf{Hyperparameters}. For all six ML4VD techniques, we used the pre-trained models and tokenizers provided by the respective authors as starting points for our experiments. Similar to Hanif et al. \cite{vulberta}, we noticed a relatively quick convergence of our performance metrics in our initial experiments on the validation dataset (after 2-7 epochs), which is why we trained each model instance for 10 epochs. For all model-specific hyperparameters, we used the values that were reported in the original papers \cite{vulberta, cotext, PLBART_paper, unixcoder_paper, codebert_paper, graphcodebert_paper}. Consult our published training scripts\footnote{GitHub: \url{https://github.com/niklasrisse/USENIX_2024}} for the complete list of hyperparameter values that we used.

\textbf{Performance metrics.} We tracked and quantified the performance of our trained models on the selected testing datasets using several commonly used performance metrics. We report six metrics in this paper: Accuracy, f1-score, precision, recall, false positive rate (FPR), and false negative rate (FNR). 

For CodeXGLUE/Devign as a balanced dataset (45.6\% vulnerable functions), we use accuracy as the main performance metric, since it is also used exclusively in the CodeXGLUE benchmark \cite{CODEXGLUE} and on the leaderboard \cite{codexglue_leaderboards}.

For VulDeePecker as a relatively imbalanced dataset (28.7\% vulnerable functions) we use the f1-score as the main performance metric. The f1-score is defined as the harmonic mean of precision and recall and is most suitable when the positive class (in our case vulnerable functions) is the minority class in an imbalanced dataset.

\textbf{Hardware}. We used a setup of five Nvidia A100 GPUs, each equipped with 40 GB RAM. One run of all our experiments takes approximately 60 days on a single A100 GPU.

\section{Experimental Results}
\label{sec:experimental_results}
\subsection*{RQ.1 Testing- and Training Data Augmentation}
We investigate, whether (a) testing data augmentation using semantic preserving transformation decreases the performance of state-of-the-art ML4VD techniques, whether (b) training data augmentation using the same transformations restores the performance towards previous levels, and whether (c) there are differences between the individual transformations.

\textbf{Methodology.}
We used \autoref{alg:methodology_1} to investigate all three questions. We ran the algorithm for each ML technique and dataset separately and measured the outcomes using the respective preferred performance metrics (see \autoref{sec:model_training_pipeline}). We did not only record the outcomes after completing the full training of the respective models but also after each training epoch in order to observe the progression of the learning process.

\textbf{Results.}
\autoref{fig:rq_1:a} shows the test set accuracy of different VulBERTa models measured after each of the ten training epochs. We can observe, that augmenting the testing dataset $Te$ with transformation $t_{10}$ leads to a substantial drop in accuracy, represented by the gap between the dotted gray and blue graphs in the figure. We can also observe, that augmenting the training dataset $Tr$ with the same transformation $t_{10}$ as the testing dataset, restores the accuracy back to previous levels (orange graph).

\begin{figure*}[t]
    \centering
    \subfloat[Test set accuracy over ten training epochs of different models trained with VulBERTa on the CodeXGLUE/Devign dataset. Augmenting the training set $Tr$ with different transformations $t_{j \neq 10}$ than the testing dataset (green lines) does not restore the accuracy back to previous levels.\label{fig:rq_2:a}]{%
      \includegraphics[width=0.47\linewidth]{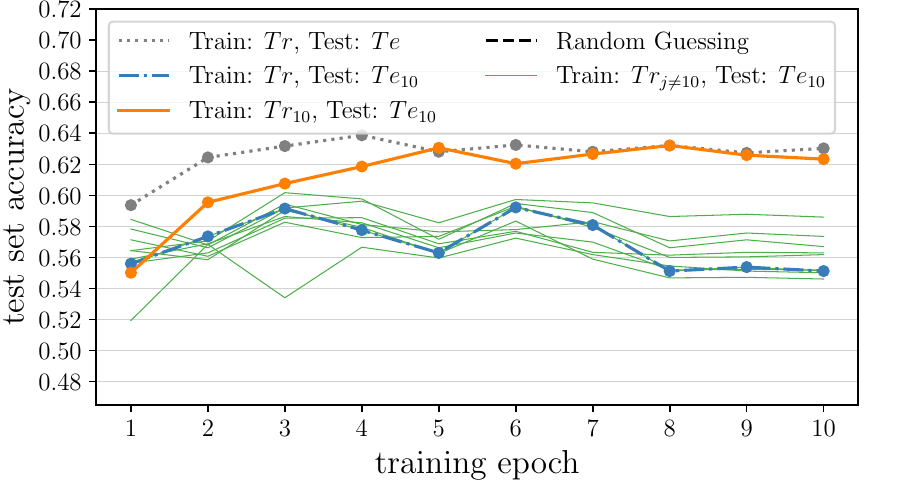}
    }
    \hfill
    \subfloat[Extension of the results in \autoref{fig:rq_2:a}, for all transformations $t_k \in T$ and for all six ML4VD techniques. The boxplots represent distributions of the resulting accuracies. Augmenting the training set $Tr$ with different transformations $t_{k \neq j}$ than the testing dataset (green boxplots) does not restore the accuracy back to previous levels. Instead of restoring, the accuracy sometimes even drops further compared to using standard training data (green below blue). \label{fig:rq_2:b}]{%
      \includegraphics[width=0.47\linewidth]{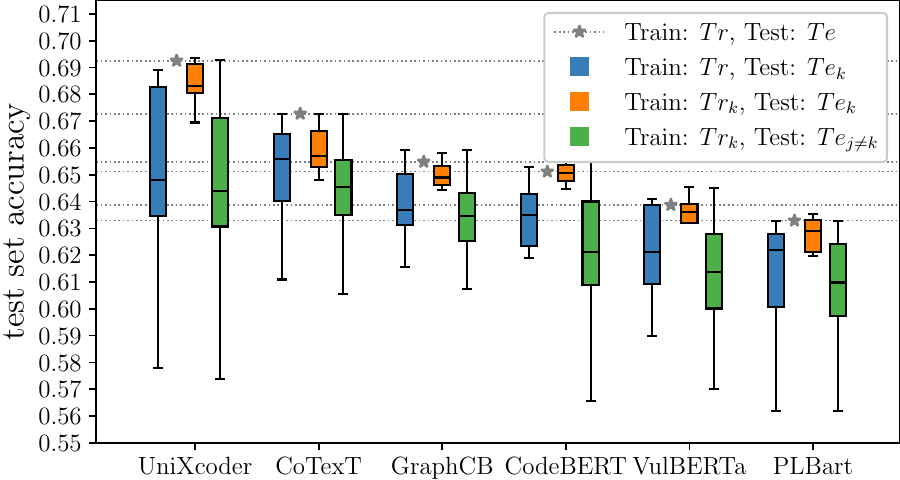}
    }
    \hfill
    \subfloat[Same setup as for \autoref{fig:rq_2:b}, but using the VulDeePecker dataset. The boxplots represent distributions of the resulting f1-scores. Augmenting the training set $Tr$ with different transformations $t_{k \neq j}$ than the testing dataset (green boxplots) does not restore the f1-score back to previous levels. \label{fig:rq_2:c}]{%
      \includegraphics[width=0.47\linewidth]{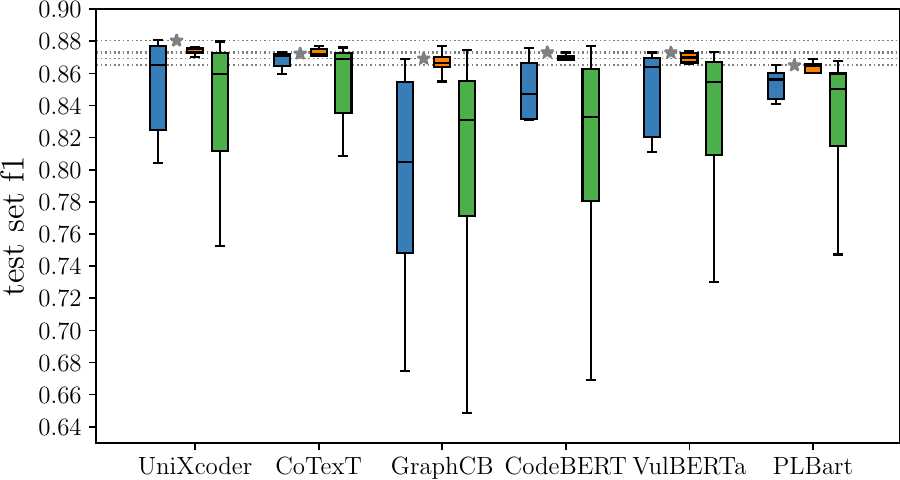}
    }
    \hfill
    \subfloat[Same setup as for \autoref{fig:rq_2:b}, but the green boxplots represent the accuracies achieved by augmenting the training data using the meta transformation $t_{11}$, in this case sampled from $\{ t_1, \ldots, t_{10} \} \setminus \{ t_j \}$, and the testing data using a single left-out transformation $t_j$. Augmenting the training dataset $Tr$ with $t_{11}$ partially restores the accuracy, although not to its previous levels. \label{fig:rq_2:d}]{%
      \includegraphics[width=0.47\linewidth]{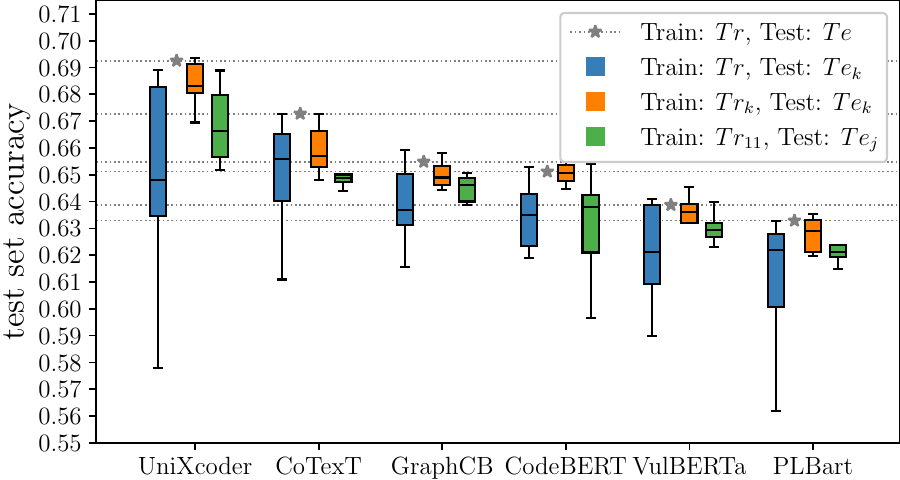}
    }
    \caption{Effects of augmenting the training data with different semantic preserving transformations than the testing data.}
    \label{fig:rq_2}
\end{figure*}

\autoref{fig:rq_1:b} extends the results of \autoref{fig:rq_1:a} to all semantic preserving transformations $t_k \in T$, and to all six ML4VD techniques. Instead of showing the accuracy for each training epoch, we only use the maximum accuracy across the full training. Across all six ML4VD techniques, we can observe, that augmenting the testing dataset $Te$ with transformations $t_k \in T$ (represented by the blue boxplots), on average, leads to a drop in accuracy compared to evaluations on the standard testing dataset $Te$ (represented by the horizontal lines with stars). In parallel to \autoref{fig:rq_1:a} we can also observe that, on average, training data augmentation using the same transformation as for testing data augmentation leads to a restoration of the observed drops in accuracy (represented by the orange boxplots).

\autoref{tab:results_alg_1} summarizes the outputs of \autoref{alg:methodology_1} for all six ML4VD techniques and both datasets. Specifically, it shows the average recorded changes in the respective metrics, when only the testing dataset was augmented (blue columns, $output_{A1.1}$), when the training and testing datasets were augmented using the same transformation (orange columns, $output_{A1.2}$), and when the training and testing datasets were augmented using a different transformation (green columns, $output_{A1.3}$). We observe that, on average, augmenting the testing dataset leads to a drop in accuracy/f1-score ($-0.025$ for CodeXGLUE, $-0.043$ for VulDeePecker), and augmenting the training dataset using the same transformation restores that decrease towards previous levels. On average, approximately 69.0\% (CodeXGLUE) and 66.2\% (VulDeePecker) of the lost accuracy/f1-score is restored.

\autoref{fig:rq_2_barchart} shows the impact on accuracy caused by augmenting only the testing data using the individual transformations $t_k \in T$ ($impact(t_k) := accuracy[MLM[Tr], Te_k] - accuracy[MLM[Tr], Te]$) for all six ML4VD techniques. In other words, \autoref{fig:rq_2_barchart} displays the severity of the performance decline of the techniques when only applying semantic preserving transformations on the testing set. The most impactful transformations for each ML technique are marked by red stars.

If a specific semantic preserving transformation has a high negative impact on the accuracy of an ML technique, we can assume that either (a) the ML technique partly relied on unrelated features that were removed or modified by the transformation (e.g. removal of comments) to achieve its original high accuracy, or (b) that the ML technique relied on unrelated features introduced by the transformation (e.g. addition of comments) to achieve the decreased accuracy after applying the transformation.

\begin{table}[t]
\centering
\scriptsize
\setlength{\tabcolsep}{1.3pt}
  \caption{Algorithm 1: Average changes when augmenting only the testing data ($output_{A1.1}$), training and testing data using the same ($output_{A1.2}$), or a different transformation ($output_{A1.3}$).}
  \label{tab:results_alg_1}
  \begin{tabular}{ll@{\hspace{3pt}}|@{\hspace{3pt}}cccc@{\hspace{3pt}}|@{\hspace{3pt}}cccc}
    \toprule
     &  & \multicolumn{4}{c|@{\hspace{3pt}}}{\bfseries CodeXGLUE} & \multicolumn{4}{c}{\bfseries VulDeePecker} \\
     \multirow{3}{*}{\rotatebox[origin=c]{90}{Metric}} &  & & \hltrafogreen{$out_{A1.1}$} & \hltrafopurple{$out_{A1.2}$} & \hltrafoyellow{$out_{A1.3}$} & & \hltrafogreen{$out_{A1.1}$} & \hltrafopurple{$out_{A1.2}$} & \hltrafoyellow{$out_{A1.3}$}\\
     &  & $Tr$ & \hltrafogreen{$Tr$} & \hltrafopurple{$Tr_k$} & \hltrafoyellow{$Tr_k$}& $Tr$ & \hltrafogreen{$Tr$} & \hltrafopurple{$Tr_k$} & \hltrafoyellow{$Tr_k$}\\
     & Technique & $Te$ & \hltrafogreen{$Te_k$} & \hltrafopurple{$Te_k$} & \hltrafoyellow{$Te_{j \neq k}$} & $Te$ & \hltrafogreen{$Te_k$} & \hltrafopurple{$Te_k$} & \hltrafoyellow{$Te_{j \neq k}$}\\
    \midrule
\multirow{6}{*}{\rotatebox[origin=c]{90}{accuracy}} & UniXcoder & 0.693 & -0.043  & -0.011  & -0.050  & 0.975 & -0.005  & -0.002  & -0.010  \\
 & CoTexT & 0.673 & -0.022  & -0.014  & -0.030  & 0.973 & -0.001  & -0.001  & -0.003  \\ 
 & GraphCB & 0.655 & -0.015  & -0.006  & -0.021  & 0.973 & -0.014  & -0.002  & -0.015  \\ 
 & CodeBERT & 0.651 & -0.034  & -0.000  & -0.040  & 0.974 & -0.007  & -0.002  & -0.012  \\ 
 & VulBERTa & 0.639 & -0.017  & -0.004  & -0.025  & 0.973 & -0.011  & -0.002  & -0.012  \\ 
 & PLBart & 0.633 & -0.021  & -0.007  & -0.026  & 0.972 & -0.003  & -0.002  & -0.008  \\ \hline
 &  &  & \textbf{-0.025}  & \textbf{-0.007}  & \textbf{-0.032}  &  & \textbf{-0.007}  & \textbf{-0.002}  & \textbf{-0.010}  \\ \hline
\multirow{6}{*}{\rotatebox[origin=c]{90}{f1-score}} & UniXcoder & 0.680 & -0.037  & -0.007  & -0.041  & 0.880 & -0.028  & -0.012  & -0.054  \\ 
 & CoTexT & 0.635 & 0.001  & 0.006  & 0.006  & 0.872 & -0.006  & -0.006  & -0.020  \\ 
 & GraphCB & 0.629 & -0.024  & -0.013  & -0.033  & 0.869 & -0.093  & -0.006  & -0.091  \\ 
 & CodeBERT & 0.596 & -0.005  & 0.012  & 0.001  & 0.873 & -0.045  & -0.007  & -0.082  \\ 
 & VulBERTa & 0.652 & -0.050  & -0.014  & -0.048  & 0.873 & -0.082  & -0.009  & -0.073  \\ 
 & PLBart & 0.618 & -0.009  & -0.006  & -0.016  & 0.865 & -0.014  & -0.007  & -0.035  \\ \hline
 &  &  & \textbf{-0.021}  & \textbf{-0.004}  & \textbf{-0.022}  &  & \textbf{-0.045}  & \textbf{-0.008}  & \textbf{-0.059}  \\ \hline
\multirow{6}{*}{\rotatebox[origin=c]{90}{recall}} & UniXcoder & 0.787 & -0.025  & -0.016  & -0.009  & 0.893 & -0.034  & -0.011  & -0.041  \\ 
 & CoTexT & 0.851 & 0.066  & 0.074  & 0.105  & 0.900 & -0.004  & -0.015  & -0.029  \\ 
 & GraphCB & 0.661 & -0.023  & -0.029  & -0.031  & 0.895 & -0.153  & -0.005  & -0.101  \\ 
 & CodeBERT & 0.572 & 0.079  & 0.031  & 0.106  & 0.890 & -0.090  & -0.004  & -0.103  \\ 
 & VulBERTa & 0.759 & -0.097  & -0.029  & -0.056  & 0.898 & -0.094  & -0.000  & -0.073  \\ 
 & PLBart & 0.658 & 0.025  & 0.003  & 0.024  & 0.884 & -0.008  & -0.003  & -0.015  \\ \hline
 &  &  & \textbf{0.004}  & \textbf{0.006}  & \textbf{0.023}  &  & \textbf{-0.064}  & \textbf{-0.006}  & \textbf{-0.060}  \\ \hline
\multirow{6}{*}{\rotatebox[origin=c]{90}{precision}} & UniXcoder & 0.717 & -0.033  & 0.020  & -0.033  & 0.939 & -0.008  & 0.030  & -0.045  \\ 
 & CoTexT & 0.684 & -0.029  & -0.027  & -0.044  & 0.955 & -0.015  & 0.013  & -0.012  \\ 
 & GraphCB & 0.734 & -0.026  & -0.019  & -0.035  & 0.995 & -0.016  & -0.005  & -0.040  \\ 
 & CodeBERT & 0.847 & -0.115  & -0.009  & -0.107  & 1.000 & -0.000  & -0.007  & -0.031  \\ 
 & VulBERTa & 0.643 & -0.004  & -0.019  & -0.034  & 0.850 & 0.023  & -0.012  & -0.010  \\ 
 & PLBart & 0.675 & -0.031  & -0.015  & -0.049  & 0.971 & -0.048  & 0.008  & -0.068  \\ \hline
 &  &  & \textbf{-0.040}  & \textbf{-0.012}  & \textbf{-0.051}  &  & \textbf{-0.011}  & \textbf{0.005}  & \textbf{-0.034}  \\ \hline
\multirow{6}{*}{\rotatebox[origin=c]{90}{FPR}} & UniXcoder & 0.142 & 0.033  & -0.018  & 0.049  & 0.006 & 0.001  & -0.003  & 0.006  \\ 
 & CoTexT & 0.211 & 0.018  & 0.028  & 0.034  & 0.004 & 0.001  & -0.001  & 0.001  \\ 
 & GraphCB & 0.079 & 0.040  & -0.003  & 0.027  & 0.000 & 0.002  & 0.000  & 0.003  \\ 
 & CodeBERT & 0.026 & 0.140  & 0.005  & 0.145  & 0.000 & 0.000  & 0.001  & 0.003  \\ 
 & VulBERTa & 0.194 & -0.036  & 0.028  & 0.040  & 0.018 & -0.004  & 0.002  & -0.001  \\ 
 & PLBart & 0.137 & 0.016  & 0.020  & 0.026  & 0.003 & 0.004  & -0.001  & 0.007  \\ \hline
 &  &  & \textbf{0.035}  & \textbf{0.010}  & \textbf{0.053}  &  & \textbf{0.001}  & \textbf{-0.000}  & \textbf{0.003}  \\ \hline
\multirow{6}{*}{\rotatebox[origin=c]{90}{FNR}} & UniXcoder & 0.213 & 0.025  & 0.016  & 0.009  & 0.107 & 0.034  & 0.011  & 0.041  \\ 
 & CoTexT & 0.149 & -0.066  & -0.074  & -0.105  & 0.100 & 0.004  & 0.015  & 0.029  \\ 
 & GraphCB & 0.339 & 0.023  & 0.029  & 0.031  & 0.105 & 0.153  & 0.005  & 0.101  \\ 
 & CodeBERT & 0.428 & -0.079  & -0.031  & -0.106  & 0.110 & 0.090  & 0.004  & 0.103  \\ 
 & VulBERTa & 0.241 & 0.097  & 0.029  & 0.056  & 0.102 & 0.094  & 0.000  & 0.073  \\ 
 & PLBart & 0.342 & -0.025  & -0.003  & -0.024  & 0.116 & 0.008  & 0.003  & 0.015  \\ \hline
 &  &  & \textbf{-0.004}  & \textbf{-0.006}  & \textbf{-0.023}  &  & \textbf{0.064}  & \textbf{0.006}  & \textbf{0.060}  \\ 
    \bottomrule
  \end{tabular}
\end{table}

From the results presented in \autoref{fig:rq_2_barchart} we can observe, that there are clear differences both between transformations and ML4VD techniques. As one might expect, a trivial transformation such as adding whitespace ($t_7$) has little to no impact on the accuracy of all six ML4VD techniques. The severity of this impact is, on average, below $1\%$ accuracy, which means that below $1\%$ of predictions are changed from correct to incorrect by applying this transformation. The ML4VD techniques also seem to be robust against identifier renaming ($t_1$ and $t_3$) and removal of comments ($t_9$), for which the severity of impact is also below or close to $1\%$. The most impactful transformations are changing the order of the function parameters ($t_2$), defining an additional void function ($t_8$), and adding code snippets from the training set as comments ($t_{10}$). For these transformations, a substantial part of the predictions is changed from correct to incorrect: Between 2.3\% and 10.6\% for $t_2$, between 1.8\% and 18.1\% for $t_8$, and between 2.1\% and 11.5\% for $t_{10}$. Overall, transformations that insert statements (e.g. $t_4$, $t_5$, $t_8$, and $t_{10}$) or reorder statements (e.g. $t_2$ and $t_6$) seem to have a higher impact than the other types.

Additionally, there are also differences between the six ML4VD techniques. For example, moving the code into a separate function ($t_6$) only seems to have a high impact on CodeBERT, and inserting a simple comment ($t_5$) seems to have a much higher impact on UniXcoder than on the other ML4VD techniques. Future work is required to determine why the ML4VD techniques are more or less robust against specific transformations.

We also investigated the impact of each individual transformation when not only the testing data but also the training data is augmented using a different transformation than for the testing data. However, due to the results being very similar to \autoref{fig:rq_2_barchart}, we decided to omit this from the paper and provide it as supplementary material in \autoref{appendix:impact_of_transformations}.

\result{Across two datasets, six ML4VD techniques, and 11 transformations, on average, (a) testing data augmentation using semantic preserving transformations leads to a drop in accuracy/f1-score (CodeXGLUE: $-0.025$, VulDeePecker: $-0.043$), (b) training data augmentation using the same transformations restores 69.0\% (CodeXGLUE) and 66.2\% (VulDeePecker) of the lost accuracy/f1-score, and (c) transformations that insert or reorder statements seem to be more impactful than other types of transformations.}

RQ.1 has already been studied in the literature, for many different techniques, datasets, and tasks \cite{AVERLOC, Generating_Adversarial_Computer_Programs_using_Optimized_Obfuscations, DAMP, ZigZag, CARROT, ALERT, Metropolis_Hastings, Abstain_refined_adv_train}. Based on our evidence, we can approve the findings of the literature.

\subsection*{RQ.2 Overfitting to Specific Transformations}
\label{sec:exp_results:rq_2}
We investigate, whether the performance of ML4VD techniques can still be restored if we augment the training dataset with a different semantic preserving transformation than the testing dataset. We use \autoref{alg:methodology_1} to investigate RQ.2, with the same setup as for RQ.1.

\textbf{Results.}
\autoref{fig:rq_2:a} is similar to \autoref{fig:rq_1:a}, it also shows the test set accuracies of different VulBERTa models measured after each of the ten training epochs. In addition to the results displayed in \autoref{fig:rq_1:a}, \autoref{fig:rq_2:a} shows the accuracies (green lines) of VulBERTa models trained on data that was augmented using all transformations except $t_{10}$, which was used to augment the testing dataset. We observe, that augmenting the training dataset $Tr$ with different transformations $t_{j \neq 10}$ as the testing dataset does not restore the accuracy back to previous levels.

\autoref{fig:rq_2:b} visualizes the same extended results for all semantic preserving transformations $t_k \in T$ and for all six ML4VD techniques. Again, the blue and the orange boxplots represent the distributions of accuracies, when either only the testing dataset (blue) or training and testing datasets were augmented using the same transformation (orange). The green boxplots represent the distribution of accuracies achieved by models that were trained on data, which was augmented using a different transformation than for the testing data. Across all six ML4VD techniques, we observe that, on average, augmenting the training dataset $Tr$ with a different transformation $t_{k \neq j}$ than the testing dataset does not restore the accuracy back to previous levels. 

\autoref{fig:rq_2:c} visualizes the same results as \autoref{fig:rq_2:b}, but using the VulDeePecker dataset. In this figure, the y-axis measures the f1-score, since it is the preferred evaluation metric for the VulDeePecker dataset. Again, we observe that, on average, augmenting the training dataset $Tr$ with a different transformation $t_{k \neq j}$ as the testing dataset does not restore the accuracy back to previous levels.

In \autoref{fig:rq_2:d} the green boxplots represent the distribution of accuracies achieved by augmenting the training data using our \emph{meta transformation} $t_{11}$. Slightly different to the definition of $t_{11}$ in \autoref{tab:transformations}, each function in the training set is transformed using a random transformation $t_k$ with $k \in [1,10] \setminus{j}$, with one left-out transformation $t_j$ which is applied to the testing data. Since our set of implemented transformations contains groups of similar transformations (e.g. adding different types of comments), we would expect the accuracies to be higher compared to applying only a single different transformation to the training set (green boxplots of \autoref{fig:rq_2:d}), but lower compared to applying exactly the same transformation to the training set (orange boxplots). Based on \autoref{fig:rq_2:d}, we observe that is the case across all six ML4VD techniques. Augmenting the training dataset $Tr$ with the meta transformation $t_{11}$ does not fully restore the accuracy, but moves it closer towards the accuracy on unaugmented data compared to applying only a single different transformation to the training set.

In addition to the results for RQ.1, \autoref{tab:results_alg_1} also shows the average recorded changes in the respective metrics, when the training and testing datasets were augmented using different transformations (green columns, $output_{A1.3}$). We observe that, on average, the score drops by 0.032 accuracy (CodeXGLUE) and 0.059 f1-score (VulDeePecker). Across the six techniques, the decrease is on average 30.2\% (CodeXGLUE) and 77.5\% (VulDeePecker) stronger than for training on unaugmented data. In other words, augmenting the training dataset using a different transformation than for the testing dataset did not restore the score towards previous levels, but instead decreased it even further. 

For the other metrics (recall, precision, FPR and FNR), we generally observe similar patterns than for accuracy and f1-score. However, there are also slight deviations, e.g. for CodeXGLUE recall improves on average by 0.023 when training data is augmented using a different transformation than the testing data instead of decreasing as expected. These deviations can be explained by innate tradeoffs between recall/precision and FPR/FNR, and can only be intepreted in context of the other metrics. Accuracy and f1-score provide a better summary of the performance, which is why they are used as the preferred metrics for the two datasets.

\result{Across two datasets, six ML4VD techniques, and 11 transformations, augmenting the training dataset using a different transformation than for the testing dataset does not restore the performance back to previous levels. In other words, the ML4VD techniques overfit to the label-unrelated features introduced by semantic preserving transformations during training data augmentation.}

\begin{table}[t]
\centering
\scriptsize
\setlength{\tabcolsep}{1.3pt}
  \caption{Algorithm 2: Performance of six ML4VD techniques evaluated on the standard CodeXGLUE/Devign testing dataset $Te$ or the vulnerability-patch testing dataset $VPTe$.}
  \label{tab:results_alg_2}
  \begin{tabular}{ll@{\hspace{3pt}}|@{\hspace{3pt}}cccc}
    \toprule
    \multirow{3}{*}{\rotatebox[origin=c]{90}{Metric}} & & $out_{A2.1}$ & \hlpurple{$out_{A2.2}$} & $out_{A2.3}$ & \hlyellow{$out_{A2.4}$} \\
    & & $Tr$ & \hlpurple{$Tr$} & $VPTr$ & \hlyellow{$VPTr$} \\
     & Technique & $Te$ & \hlpurple{$VPTe$} & Test: $VPTe$ & \hlyellow{$Te$} \\
    \midrule
    \multirow{6}{*}{\rotatebox[origin=c]{90}{accuracy}} & UniXcoder & 0.693 & 0.414 & 0.616 & 0.546 \\ 
     & CoTexT & 0.673 & 0.503 & 0.607 & 0.575 \\ 
     & GraphCB & 0.655 & 0.342 & 0.596 & 0.546 \\ 
     & CodeBERT & 0.651 & 0.294 & 0.571 & 0.548 \\ 
     & VulBERTa & 0.639 & 0.527 & 0.602 & 0.564 \\ 
     & PLBart & 0.633 & 0.524 & 0.598 & 0.572 \\ \hline
     & & \textbf{0.657} & \textbf{0.434} & \textbf{0.598} & \textbf{0.559}\\ \hline
    \multirow{6}{*}{\rotatebox[origin=c]{90}{f1-score}} & UniXcoder & 0.680 & 0.582 & 0.662 & 0.613 \\ 
     & CoTexT & 0.635 & 0.667 & 0.665 & 0.616 \\ 
     & GraphCB & 0.629 & 0.508 & 0.654 & 0.603 \\ 
     & CodeBERT & 0.596 & 0.455 & 0.629 & 0.613 \\ 
     & VulBERTa & 0.652 & 0.610 & 0.651 & 0.615 \\ 
     & PLBart & 0.618 & 0.583 & 0.633 & 0.575 \\ \hline
     & & \textbf{0.635} & \textbf{0.567} & \textbf{0.649} & \textbf{0.606}\\ \hline
    \multirow{6}{*}{\rotatebox[origin=c]{90}{recall}} & UniXcoder & 0.787 & 0.819 & 0.870 & 0.896 \\ 
     & CoTexT & 0.851 & 1.000 & 0.975 & 0.941 \\ 
     & GraphCB & 0.661 & 0.680 & 0.835 & 0.873 \\ 
     & CodeBERT & 0.572 & 0.589 & 0.770 & 0.883 \\ 
     & VulBERTa & 0.759 & 0.758 & 0.909 & 0.928 \\ 
     & PLBart & 0.658 & 0.680 & 0.741 & 0.738 \\ \hline
     & & \textbf{0.715} & \textbf{0.754} & \textbf{0.850} & \textbf{0.876}\\ \hline
    \multirow{6}{*}{\rotatebox[origin=c]{90}{precision}} & UniXcoder & 0.717 & 0.452 & 0.668 & 0.518 \\ 
     & CoTexT & 0.684 & 0.502 & 0.724 & 0.702 \\ 
     & GraphCB & 0.734 & 0.406 & 0.622 & 0.509 \\ 
     & CodeBERT & 0.847 & 0.371 & 0.656 & 0.516 \\ 
     & VulBERTa & 0.643 & 0.531 & 0.781 & 0.647 \\ 
     & PLBart & 0.675 & 0.535 & 0.663 & 0.547 \\ \hline
     & & \textbf{0.717} & \textbf{0.466} & \textbf{0.686} & \textbf{0.573}\\ \hline
    \multirow{6}{*}{\rotatebox[origin=c]{90}{FPR}} & UniXcoder & 0.142 & 0.816 & 0.107 & 0.172 \\ 
     & CoTexT & 0.211 & 0.823 & 0.060 & 0.041 \\ 
     & GraphCB & 0.079 & 0.840 & 0.091 & 0.177 \\ 
     & CodeBERT & 0.026 & 0.849 & 0.102 & 0.233 \\ 
     & VulBERTa & 0.194 & 0.312 & 0.034 & 0.061 \\ 
     & PLBart & 0.137 & 0.251 & 0.138 & 0.213 \\ \hline
     & & \textbf{0.131} & \textbf{0.649} & \textbf{0.089} & \textbf{0.149}\\ \hline
    \multirow{6}{*}{\rotatebox[origin=c]{90}{FNR}} & UniXcoder & 0.213 & 0.181 & 0.130 & 0.104 \\ 
     & CoTexT & 0.149 & 0.000 & 0.025 & 0.059 \\ 
     & GraphCB & 0.339 & 0.320 & 0.165 & 0.127 \\ 
     & CodeBERT & 0.428 & 0.411 & 0.230 & 0.117 \\ 
     & VulBERTa & 0.241 & 0.242 & 0.091 & 0.072 \\ 
     & PLBart & 0.342 & 0.320 & 0.259 & 0.262 \\ \hline
     & & \textbf{0.285} & \textbf{0.246} & \textbf{0.150} & \textbf{0.124}\\
    \bottomrule
  \end{tabular}
\end{table}

In summary, we can observe that across the tested ML4VD techniques, transformations, and datasets, training data augmentation only restores the performance to previous levels when the testing dataset is augmented in a similar way than the training dataset.

\result{The performance gained by data augmentation only applies to the specific transformations used during the training of the model. ML4VD techniques continue to leverage unrelated features when deciding whether a function contains a security vulnerability. }

\subsection*{RQ.3 Generalization to VulnPatchPairs}
\label{sec:exp_results:rq_3}
We investigate, whether (a) ML4VD techniques are able to generalize from typical vulnerability detection training datasets to a modified setting, in which they are required to distinguish between vulnerabilities and their patches, and (b) whether training to distinguish between vulnerabilities and patches improves the performance on standard testing data.

\textbf{Methodology.} We used \autoref{alg:methodology_2} to investigate both questions. As inputs to the algorithm, we selected the training and testing subsets of the CodeXGLUE/Devign dataset as the standard training and testing datasets $Tr$ and $Te$, and the training and testing subsets of VulnPatchPairs as the vulnerability-patch training and testing datasets $VPTr$ and $VPTe$. We ran the algorithm for all six ML4VD techniques separately.

\textbf{Results.}
\autoref{tab:results_alg_2} shows the results of running \autoref{alg:methodology_2}. Specifically, it shows the performance of different models evaluated on the standard CodeXGLUE/Devign testing dataset $Te$ or the vulnerability-patch testing dataset $VPTe$. We focus our analysis on the results measured in accuracy since it is the preferred performance metric for balanced datasets such as CodeXGLUE and VulnPatchPairs.

We observe, that the accuracy of all six ML4VD techniques is highest (between 0.633 and 0.693) when trained and evaluated on standard training and testing data (second column, $output_{A2.1}$). This is expected and consistent with the findings in the literature \cite{vulberta, cotext, codebert_paper, graphcodebert_paper, unixcoder_paper, PLBART_paper}. When trained and evaluated on VulnPatchPairs (fourth column, $output_{A2.3}$) the accuracy is consistently lower than in the standard setting (between 0.558 and 0.617), but still significantly higher than the expected accuracy of a random guesser\footnote{Since $VPTe$ is perfectly class balanced (50\% vulnerable, 50\% clean), a random guesser (coin flip) would be expected to achieve an accuracy of 0.5.}. However, when trained on standard training data and evaluated on the VulnPatchPairs testing dataset (third column, $output_{A2.2}$), the accuracy drops dramatically (between 0.294 and 0.527). Even the best model (VulBERTa) is only 0.027 points better than a random guesser. On average, the accuracy is worse than random guessing. In other words, all six ML4VD techniques that we evaluated are unable to distinguish between vulnerabilities and their patches when trained on a typical vulnerability detection dataset. 

When trained on VulnPatchPairs and evaluated on standard testing data (fifth column, $output_{A2.4}$), we get a similar picture. The performance is significantly worse (between 0.546 and 0.575) compared to models trained on standard training data (second column). However, the performance in this case is notably better than random guessing.  

\result{(a) All six ML4VD techniques are not able to distinguish between vulnerabilities and their patches when trained on standard training data. On average, the accuracy is lower than the expected accuracy of a random guesser. (b) When trained to distinguish between vulnerabilities and their patches, the ML4VD techniques are able to predict standard testing data better than a random guesser, but still significantly worse than when trained on standard training data. In other words, the ML4VD techniques are unable to generalize from their training data to a slightly modified vulnerability detection setting.}

\section{Threats to Validity}
\label{sec:threats_to_validity}
As for any empirical study, there are various threats to the validity of our results and conclusions.

\textbf{Internal validity.} A common source of systematic error in empirical studies on ML4VD techniques is hyperparameter selection. Given a particular desired outcome, hyperparameters can be optimized to move the result in the desired direction. We tried to minimize this risk by taking the values for hyperparameters provided by the authors of the chosen ML4VD techniques.

Another potential source of systematic error is the training-/testing dataset split. Similar to hyperparameter selection, dataset split can also be varied to change a result in a desired direction. We tried to avoid this risk by taking the provided splits of the CodeXGLUE benchmark \cite{CODEXGLUE} and by Hanif et al. \cite{vulberta_vuldeepecker_dataset}.

\textbf{External validity.} The degree to which our results generalize to other learning-based techniques, vulnerability detection datasets, semantic preserving transformations, and performance metrics, are concerns of external validity. We tried to minimize the risk attached to these concerns by evaluating a wide set of six state-of-the-art techniques, two datasets, six performance metrics, and 11 semantic preserving transformations. For RQ.3, we only investigate the generalization between CodeXGLUE/Devign and VulnPatchPairs. To maximize generality, we tried to keep both \autoref{alg:methodology_1} and \autoref{alg:methodology_2} as general as possible, so that they can easily be adapted to other techniques, datasets, transformations, and metrics.

\textbf{Simplicity of Transformations.} Some of the semantic preserving transformations that we used (see \autoref{tab:transformations}) could be easily addressed by adding additional data pre-processing (e.g. mapping identifiers to standardized names). However, the specific transformations that we implemented are merely a tool to demonstrate, that the performance gained by training data augmentation only applies to the specific transformations used for training and that the techniques that we investigated overfit to the label-unrelated features introduced by these transformations. For a new technique, they could be replaced by a different set of transformations.

\textbf{Class balance.} Multiple works have shown that learning-based vulnerability detection techniques trained on fairly balanced datasets (such as CodeXGLUE) often fail to generalize to real-world code repositories \cite{distribution-shift, arewethereyet, dosanddonts}, which usually contain a much smaller ratio of security vulnerabilities \cite{frequence_of_bugs}. However, measured by citations, class-balanced datasets are still by far the most popular datasets to evaluate learning-based techniques for vulnerability detection. To our current knowledge, there is no vulnerability detection dataset with sufficient size (more than 10k code snippets), high-quality labels (manually provided by security experts), and a realistic distribution of vulnerable to non-vulnerable code snippets that is widely used in the research community (at least 50 citations). This is why we decided to focus our experiments on the CodeXGLUE and VulDeePecker datasets, even though they do not reflect a realistic class distribution.

\section{Discussion and Future Work}
\label{sec:discussion_and_future_work}
In summary, our results demonstrate that state-of-the-art ML4VD techniques overfit to the label-unrelated features that are introduced by semantic preserving transformations during training and that ML4VD techniques are not able to generalize to a modified setting, in which they have to distinguish between vulnerabilities and their patches.

\textbf{Overfitting} of learned models is a well-known problem in the machine learning research field \cite{overfitting_overview, overfitting_1995}. However, as shown in our experiments, the traditional approach to evaluating ML4VD techniques often fails to detect overfitting to label-unrelated features in the training data. Our proposed \autoref{alg:methodology_1} is a novel way to measure overfitting of ML4VD techniques, that goes beyond the traditional approach, and can even detect overfitting if there is no gap in the standard setup at all. There are several common strategies to reduce overfitting in the standard evaluation setup, e.g. early-stopping, dropout, or large pre-training datasets \cite{overfitting_overview}, which are already integrated in the ML4VD techniques that we used in our experiments. However, our experiments demonstrate that the techniques are still severely overfitting to label-unrelated features introduced by semantic preserving transformations during training data augmentation. Finding ways to robustify ML4VD techniques without or with minimal overfitting will be a central challenge of the machine learning for vulnerability detection research area. We hope that our proposed algorithms can be used to understand the problem, to develop new approaches, and to track the progress in this direction.

\textbf{Generalization.} The results for RQ.3 (see \autoref{sec:exp_results:rq_3}) reveal, that state-of-the-art ML4VD techniques lack the ability to generalize from their training data to a modified setting, which requires to distinguish between vulnerabilities and their patches. Since we can not assume that real-world software systems would be similar to the training data of these techniques, the ability to generalize to modified settings would be required for these techniques to be safely integrated into real software engineering environments.

The ability of a ML technique to generalize to testing data that is differently distributed than the training data is also called \emph{out-of-distribution generalization}, and the lack of it for learning-based techniques has been recently identified (e.g. in the computer vision domain \cite{ood_cv, ood_survey}). Our proposed \autoref{alg:methodology_2} can be seen as a tool to measure out-of-distribution generalization for the domain of automatic vulnerability detection. It would be an interesting avenue for future work to try approaches that have been used to address out-of-distribution generalization in other domains (e.g. causal representation learning \cite{Scholkopfetal21}) on the task of automatic vulnerability detection and measure the success using our \autoref{alg:methodology_2}.\newpage

\bibliographystyle{plain}
\bibliography{main}

\appendix
\begin{figure}[t]
    \includegraphics[width=\linewidth]{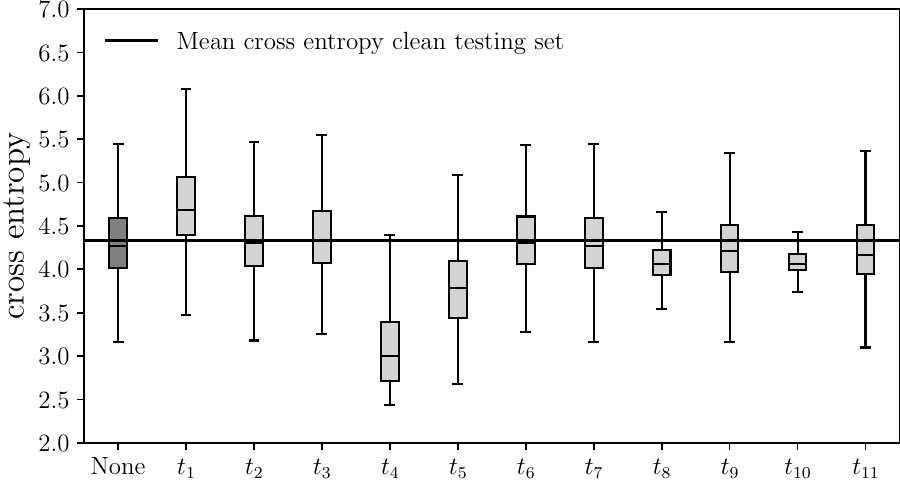}
    \caption{Naturalness of the semantic preserving transformations, that we used in our experiments. Lower cross entropy means higher naturalness. All transformations except $t_1$ (identifier renaming) lead to lower or equal cross entropy than no transformation (None). }
    \label{fig:naturalness}
\end{figure}
\begin{figure*}[t!]
    \centering
    \includegraphics[width=\linewidth]{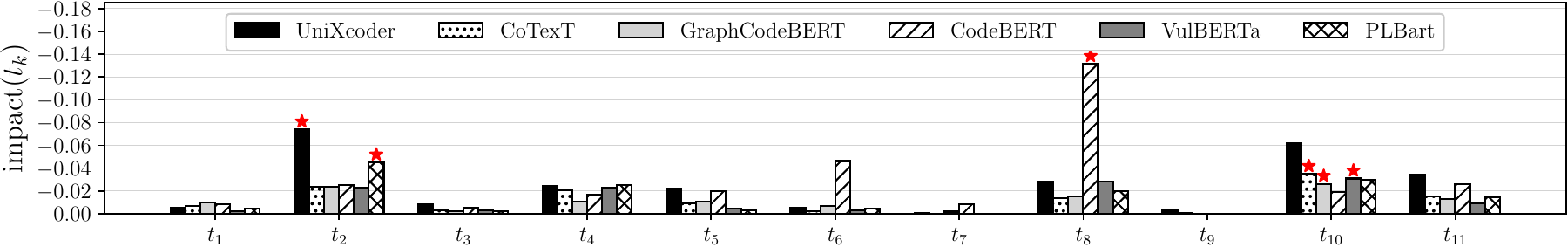}
    \caption{Impact on accuracy caused by augmenting the testing data with a different transformation than the training data ($impact(t_k) := \frac{1}{(N-1)} \sum_{t_j \in T} accuracy[MLM[Tr_j], Te] - accuracy[MLM[Tr_j], Te_k]$). The most impactful transformations for each ML technique are marked by red stars.}
    \label{fig:rq_2_barchart_appendix}
\end{figure*}
\section{Naturalness}
\label{appendix:naturalness}
\begin{figure*}[t!]
    \centering
    \subfloat[Accuracy]{%
      \includegraphics[width=0.09\linewidth]{figures/fig_rq2_boxplot_accuracy.pdf}
    }
    \hfill
    \subfloat[F1-Score]{%
      \includegraphics[width=0.09\linewidth]{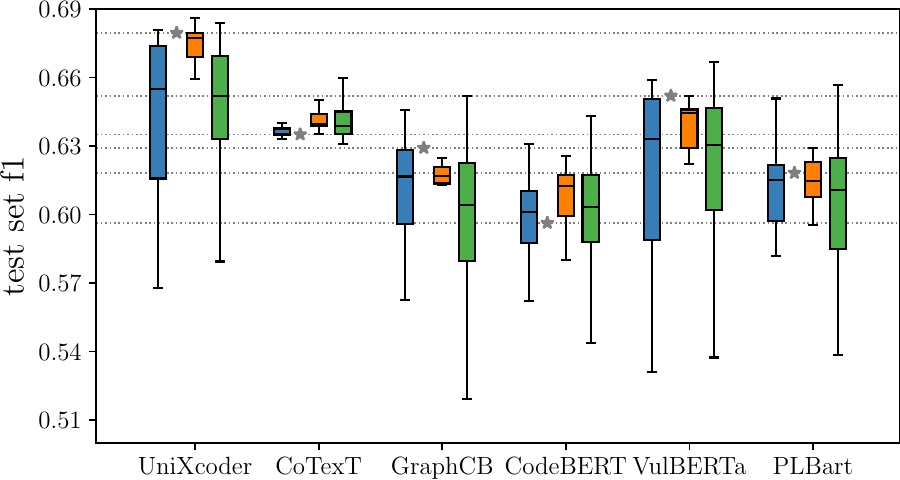}
    }
    \hfill
    \subfloat[Precision]{%
      \includegraphics[width=0.09\linewidth]{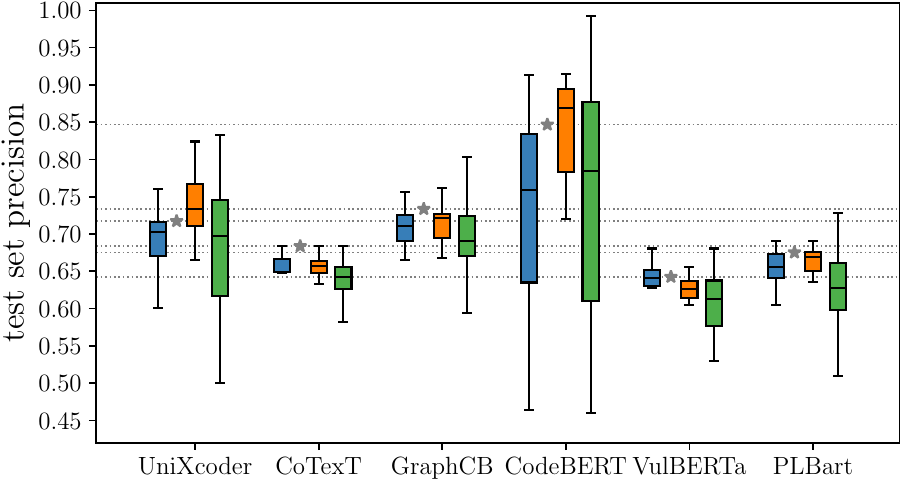}
    }
    \hfill
    \subfloat[Recall]{%
      \includegraphics[width=0.09\linewidth]{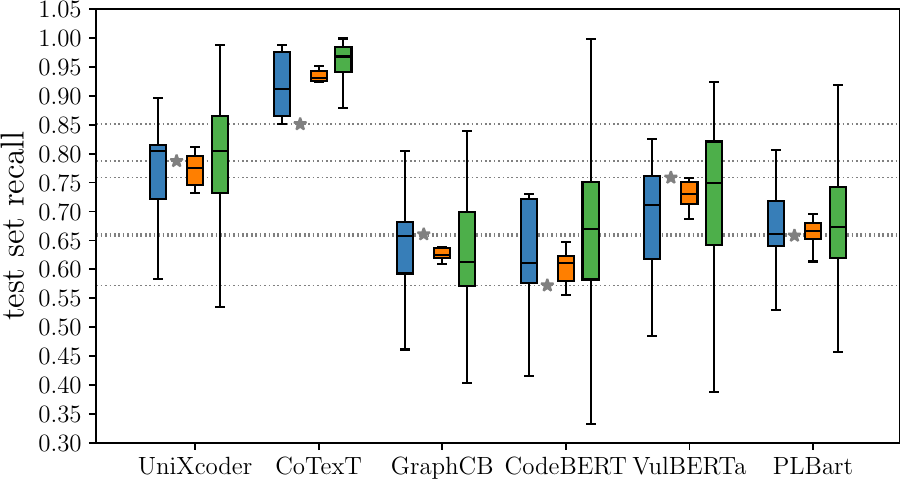}
    }
    \hfill
    \subfloat[FPR]{%
      \includegraphics[width=0.09\linewidth]{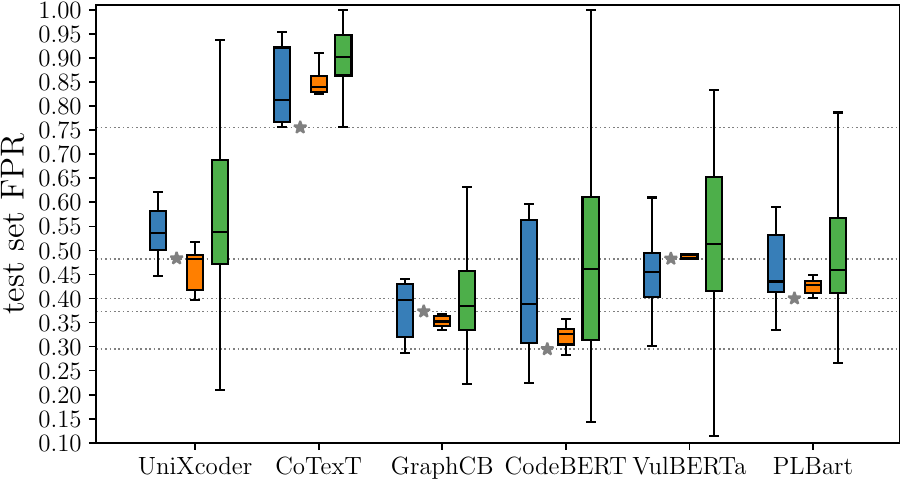}
    }
    \hfill
    \subfloat[FNR]{%
      \includegraphics[width=0.09\linewidth]{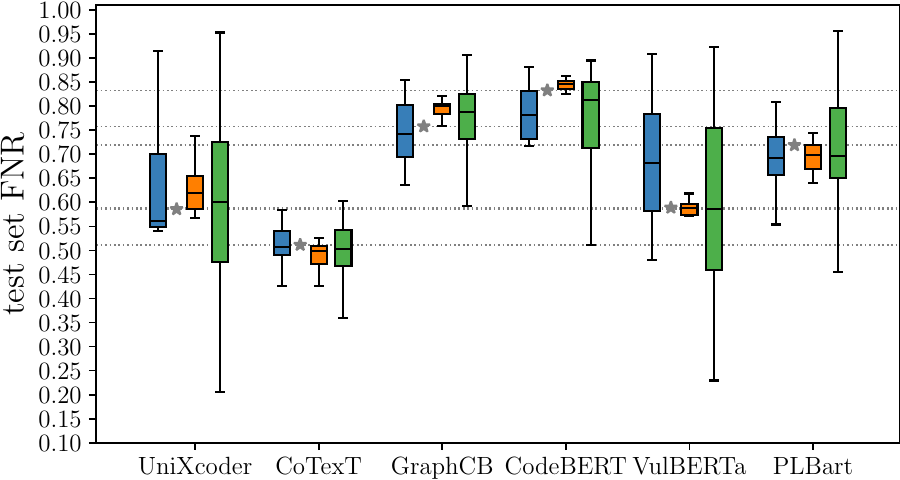}
    }
    \caption{Additional metrics for RQ.2 using the CodeXGLUE/Devign dataset. The results support the conclusions that we generated based on the main metric (accuracy).}
    \label{fig:AD:codexglue}
\end{figure*}
\begin{figure*}[t!]
    \centering
    \subfloat[Accuracy]{%
      \includegraphics[width=0.09\linewidth]{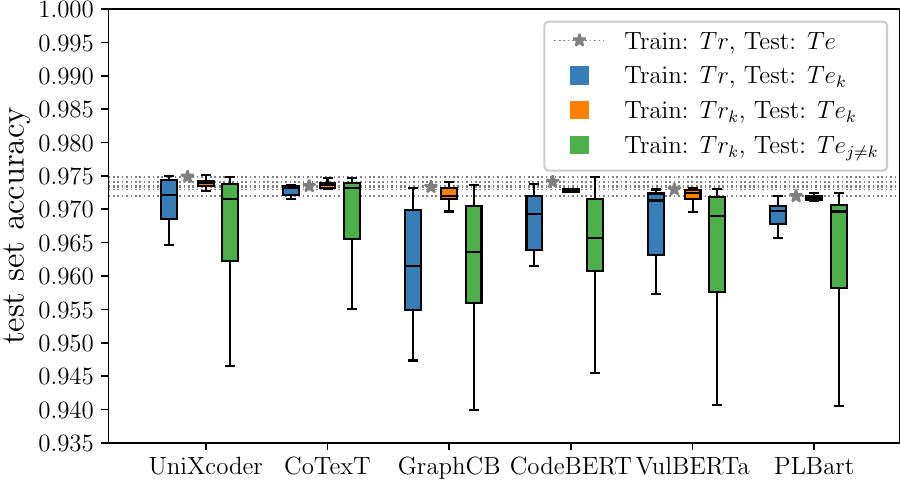}
    }
    \hfill
    \subfloat[F1-Score]{%
      \includegraphics[width=0.09\linewidth]{figures/fig_rq2_boxplot_vuldeepecker_f1.pdf}
    }
    \hfill
    \subfloat[Precision]{%
      \includegraphics[width=0.09\linewidth]{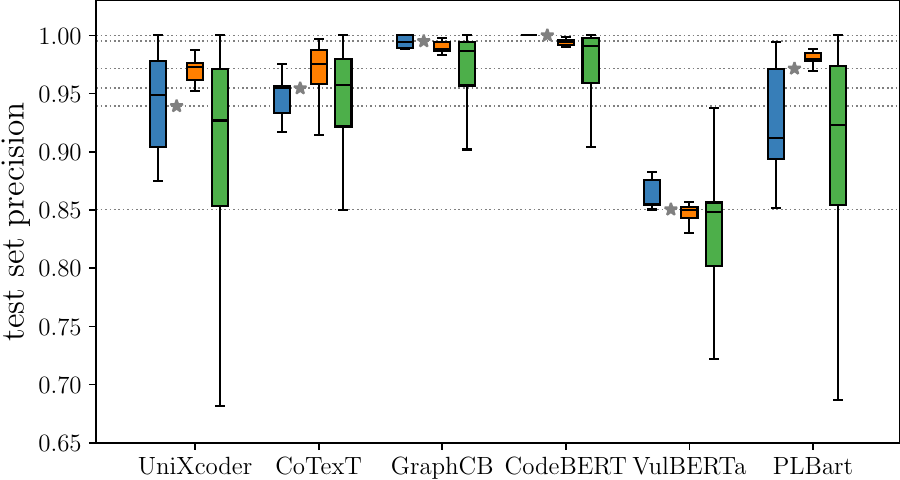}
    }
    \hfill
    \subfloat[Recall]{%
      \includegraphics[width=0.09\linewidth]{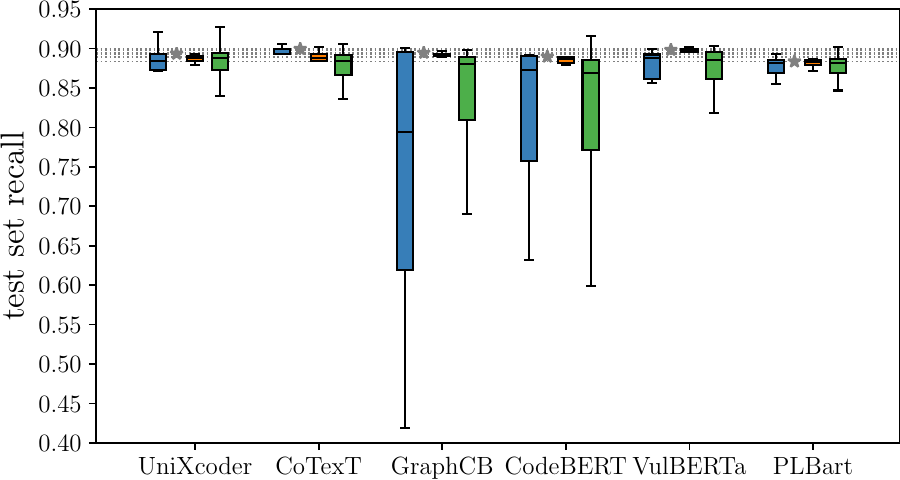}
    }
    \hfill
    \subfloat[FPR]{%
      \includegraphics[width=0.09\linewidth]{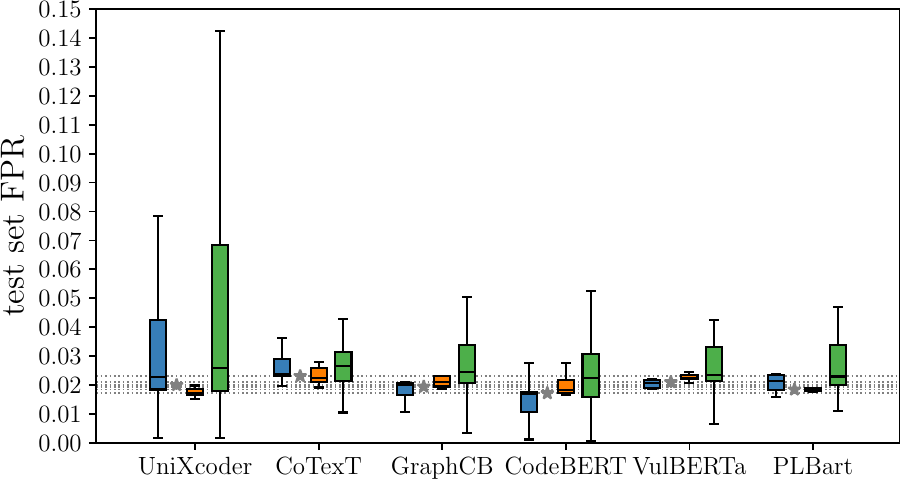}
    }
    \hfill
    \subfloat[FNR]{%
      \includegraphics[width=0.09\linewidth]{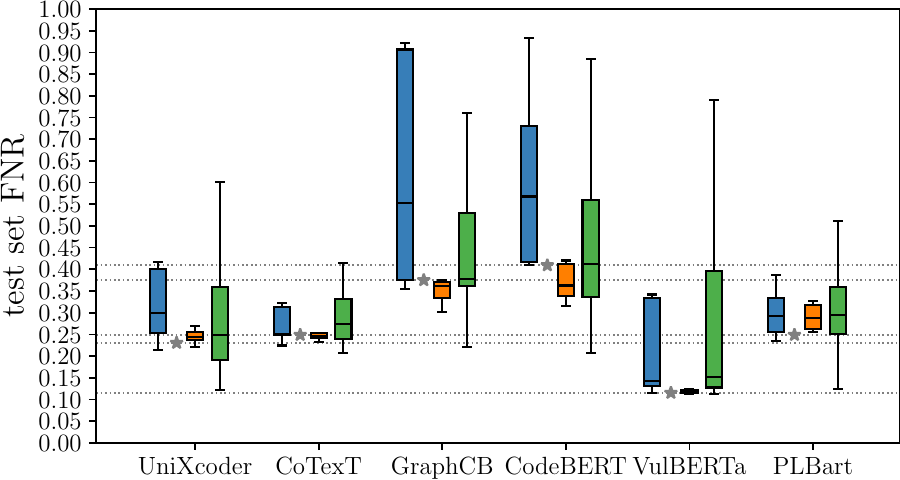}
    }
    \caption{Additional metrics for RQ.2 using the VulDeePecker dataset. The results support the conclusions that we generated based on the main metric (f1-score).}
    \label{fig:AD:vuldeepecker}
\end{figure*}
Since vulnerability detection techniques are ultimately designed to be applied to real-world code, we also need to ensure that our transformations lead to code snippets that could occur in the real world, or i.e., lead to \emph{natural} code. We measured the naturalness of our transformations using the method introduced by Hindle et al. \cite{naturalness} (implemented as a 2-gram markov model) and present the results in \autoref{fig:naturalness}. Using the method introduced by Hindle et al., we can compute the \emph{cross entropy} of a given code snippet, which represents how surprising or \emph{unnatural} the code snippet is relative to the code snippets observed in the training dataset (for a detailed explanation consult the work of Hindle et al. \cite{naturalness}). Using this approach, we computed the cross entropy for all code snippets in the CodeXGLUE testing dataset (dark gray boxplot), and for transformed versions of the CodeXGLUE testing dataset (light gray boxplots). The horizontal black line represents the average cross entropy for all code snippets in the untransformed CodeXGLUE testing dataset. We can observe, that for all transformations except $t_1$ (identifier renaming), the cross entropy is similar or lower than for the untransformed dataset. In other words, all transformations except $t_1$ (identifier renaming) lead to code snippets that are similar in naturalness compared to the real-world code of the CodeXGLUE testing dataset.
\section{Impact of Individual Transformations}
\label{appendix:impact_of_transformations}
\autoref{fig:rq_2_barchart_appendix} shows the impact on accuracy caused by augmenting the testing data with a different transformation than the training data ($impact(t_k) := \frac{1}{(N-1)} \sum_{t_j \in T} accuracy[MLM[Tr_j], Te] - accuracy[LLM[Tr_j], Te_k]$). The most impactful transformations for each LLM are marked by red stars. The results are very similar to \autoref{fig:rq_2_barchart}, which is why we chose to omit \autoref{fig:rq_2_barchart_appendix} from the main paper and provide it as supplementary material.
\section{Model Architecture Details}
\label{appendix:model_details}
All selected ML4VD techniques happen to be token-based large language models (LLMs), specialized for the task of vulnerability detection. LLMs are based on the transformer architecture, which is a neural network model architecture for sequence-to-sequence tasks based on the attention mechanism \cite{NIPS2017_3f5ee243}. The attention mechanism is essentially a weighted dot product that allows models to focus on specific parts of the input data that are most relevant to the task at hand, improving their ability to capture dependencies and context. In a transformer model, the attention mechanism is combined with parametrized feed-forward layers and repeated multiple times, resulting in a complex multi-layer network. A detailed description of the transformer architecture can be found in the original paper \cite{NIPS2017_3f5ee243}. While the base architecture is the same for all of the six techniques, there are some notable differences in the pre-training setup, size, or specific parts of the model training: 

\textbf{VulBERTa.} VulBERTa leverages a custom tokenization strategy, which is based on the \emph{byte pair encoding} algorithm \cite{byte_pair_encoding} combined with a set of pre-defined code tokens (standard C/C++ keywords, punctuation, and library API calls) to achieve better code encodings through maintaining the syntactical structure of source code.

\textbf{CoTexT.} CoTexT uses multi-task learning for pre-training, which means that the model is trained to perform multiple code-related tasks (e.g. vulnerability detection, code summarization, and code generation) in parallel. The model architecture also contains significantly more trainable parameters than the other techniques (222M parameters), because it contains more more multi-head attention layers. CoTexT also uses a different tokenizer than the other techniques.

\textbf{UniXcoder.} UniXcoder introduces a new pre-training setup, which includes tokenization of the abstract syntax trees (ASTs) of the code, and three different 'training modes' that leverage different self-attention masks.

\textbf{PLBart.} PLBart uses a special pre-training procedure called 'denoising autoencoding', which is a combination of token masking, token deletion, and token infilling, that has to be reversed by the model.

\textbf{CodeBERT.} CodeBERT uses a pre-training setup in which natural language (e.g. documentation) and code are combined to produce a more semantically stable representation of the input.

\textbf{GraphCodeBERT.} GraphCodeBERT uses a pre-training task, where in addition to natural language and the code, a graph-based representation of the data flow of the function is provided.
\section{Additional Metrics for RQ.2}
\label{appendix:additional_metrics}
\autoref{fig:AD:codexglue} and \autoref{fig:AD:vuldeepecker} show additional results for RQ.2 using all of our available metrics (accuracy, f1-score, precision, recall, FPR, and FNR). Generally, the results support the conclusions that we generated based on the respective preferred metrics on the two datasets (accuracy for CodeXGLUE/Devign and f1-score for VulDeePecker). While the observed patterns deviate for some of the metrics (e.g. precision, recall, FPR, or FNR), these deviations can be explained by the relationships between them. For example, a model can have a really high precision, but low recall. Similarly, a model can have really low false-negative-rate, but the corresponding false-positive-rate is really high. Accuracy and f1-score provide a better summary of the performance, which is why they are used as the preferred metrics for the two datasets.

\end{document}